\documentclass[a4paper, 12pt]{article}
\usepackage{inputenc}
\usepackage{amsfonts}
\usepackage{amsmath}
\usepackage{amssymb}
\usepackage{amsthm}
\usepackage{bbm}
\usepackage{booktabs}
\usepackage{float}
\usepackage{graphicx} 
\usepackage{graphics}
\usepackage[semicolon,authoryear]{natbib}
\usepackage[dvipsnames]{xcolor}
\usepackage{comment}
\usepackage{tikz}
\usetikzlibrary{positioning}
\usepackage{hyperref}
\usepackage[noabbrev,capitalize]{cleveref}
\usepackage{marvosym}
\usepackage{subcaption}

\definecolor{ViridisGreen}{HTML}{009E73}
\definecolor{ViridisBlue}{HTML}{0072B2}
\definecolor{ViridisOrange}{HTML}{D55E00}
\newcommand{\TT}{\mathbf{T}}
\newcommand{\QQ}{\mathbf{Q}}
\newcommand{\HH}{\mathbf{H}}

\title{Model-based Clustering of Compositional Trajectories for the Analysis of Mobility Data}

\author{Andrea Panarotto$^{1,*}$,
Manuela Cattelan$^{1}$, and Ruggero Bellio$^{2}$\\
\small
$^{1}$Department of Statistical Sciences, University of Padova, Padova, Italy\\
\small
$^{2}$Department of Economics and Statistics, University of Udine, Udine, Italy\\
\small
\Letter andrea.panarotto@unipd.it
	   }
\date{}
\begin{document}
\maketitle
\begin{abstract}
    Understanding urban mobility patterns is crucial for designing efficient and sustainable transportation systems. Motivated by an application to the municipality of Padova and its surroundings, we propose a novel statistical framework for the analysis and clustering of mobility trajectories derived from telephonic data. We introduce a compositional representation of individual movements that integrates the uncertain device location with information on the surrounding road network, encoding at each time point the proportions of different road types compatible with the observed position. This formulation naturally accounts for measurement uncertainty and yields trajectories evolving in the simplex.
    To model these data, we develop a state-space framework for compositional time series that captures both the telephonic measurement error and the temporal dynamics of the latent mobility process. Building on this representation, we propose a model-based clustering approach based on mixtures of state-space models to identify groups of trajectories with similar evolution. This allows us to aggregate individual movements into interpretable mobility patterns at the population level.
    The results of the case study demonstrate the ability of the approach to uncover meaningful mobility behaviors, providing insights that are potentially relevant to policy makers.
\end{abstract}

\section{Introduction}
\label{sec:intro}
The study of urban mobility is becoming increasingly important in understanding the functioning of modern cities. Assessing how transportation infrastructures are used and identifying the underlying patterns of mobility are essential for effective transportation planning and management. Uncovering recurrent mobility behaviors allows researchers and policy makers to better understand the interaction between users and the transportation network, providing evidence to address questions related to infrastructure use, traffic organization, and mobility demand. These insights can also support the development of future sustainable transportation policies.

Telephonic data represent a powerful tool for investigating such patterns. Collected by telecommunication companies for operational purposes, they generate large-scale datasets that can be repurposed to study human mobility. Sources such as call detail records (CDRs), location updates, and signaling data provide information at increasingly fine temporal and spatial resolutions, offering unprecedented opportunities to observe population movements. Early studies were mainly based on aggregated CDRs, enabling the estimation of population-level quantities such as the radius of gyration, step lengths, or waiting times between consecutive displacements \citep{gonzalez2008understanding,song2010modelling,zhao_explaining_2015}, as well as the reconstruction of origin–destination matrices and traffic flows using descriptive approaches or heuristic algorithms \citep{white_extracting_2002,calabrese2011estimating,wang_understanding_2012,iqbal2014development,alexander2015origin,ccolak2015analyzing,toole2015path,jiang2017activity}. As temporal resolution improved, research progressively shifted toward the individual scale, leading to the development of statistical models for activity detection \citep{liao2005location,eagle2006reality,widhalm2015discovering,yin2017generative,vana2022posterior} and recognition of transportation modes \citep{wang2010transportation,xu_transportation_2011,kalatian_travel_2016,graells-garrido_inferring_2018,bachir2019inferring,huang2019transport}. In parallel, generative models have been proposed to mitigate privacy concerns and compensate for incomplete observations by simulating realistic mobility patterns \citep{song2010modelling,jiang2016timegeo,pappalardo2016human,pappalardo2018data,pappalardo2022scikit,feng2020learning,yin2017generative,wang_large_2021,berke2022generating,luca_survey_2023}.

Despite this rich body of work, contributions from the statistical literature remain relatively limited, as the field has been predominantly driven by transport engineering and computer science. We argue that statistical methodology can play a crucial role in advancing mobility research. The complexity and volume of telephonic data are matched by substantial inferential challenges, especially when dealing with tower-based localization technologies, where recorded positions are affected by non-negligible spatial uncertainty. Properly accounting for this intrinsic measurement error requires principled probabilistic modeling and careful treatment of uncertainty, which naturally fall within the domain of statistics.

To address these challenges, we propose a novel framework for the analysis of urban mobility based on compositional data analysis \citep{aitchison1982statistical,aitchison1986statistical}. To our knowledge, this is the first application of compositional methods in this context, although such techniques are well established in fields such as microbiology, geology, social sciences, and finance \citep[e.g.][]{greenacre2023aitchison}. We represent individual trajectories as compositional time series, where each observation lies in the simplex and encodes the proportions of different types of roads in the individual’s surroundings at a given time. This representation naturally captures the relative nature and inherent dependence structure of these quantities, allowing us to model mobility patterns while explicitly accounting for their compositional constraints.

The analysis of compositional time series remains relatively underexplored in the literature and requires dedicated methodological tools \citep{grunwald_time_1993,bruno_clustering_2011,kynclova_modeling_2015,zheng_dirichlet_2017}. We introduce a state-space model tailored to compositional data that simultaneously accounts for telephonic measurement uncertainty and for the temporal evolution of the latent compositional trajectory. Building on this framework, we develop a model-based clustering approach for the unsupervised grouping of compositional trajectories through a mixture of state-space models. Clustering these trajectories is our ultimate goal, as it enables us to recover interpretable mobility patterns and to provide insight related to the issues mentioned at the outset, including differences in travel modes, traveler typologies, and contrasts between working days and holidays.

The remainder of the paper is organized as follows. \cref{sec:data_description} presents a detailed description of the data and introduces the proposed compositional representation of mobility trajectories. \cref{sec:model_definition} describes the mixture of state-space models used for model-based clustering. 
\cref{sec:application} is devoted to the motivating application to urban mobility in the city of Padova (Italy), which represents the central contribution of the paper, illustrating how the proposed framework translates into practically relevant insights for the analysis of real mobility data. \cref{sec:discussionState} concludes with final remarks and a discussion, while \cref{sec:data_access} provides additional information on data access and privacy constraints related to the use of sensitive telephonic data.

\section{Compositional representation of telephonic mobility data}
\label{sec:data_description}

Urban mobility trajectories are derived from mobile phone signals through tower-mounted amplifier logs. Mobile phones continuously attempt to maintain a connection to the cellular network by sending numerous pings per minute to nearby telecommunication towers. Each tower provides coverage to its vicinity through multiple cells, which emit wireless signals at various frequencies and in various directions. It is possible to estimate the coverage area of each individual cell and determine to which cell a mobile phone is connected at any given moment; see \url{www.opencellid.org} for details. Consequently, when a mobile phone connects to a cell, the center of mass of the cell's coverage area serves as a proxy for the user's location. A trajectory is the sequence of these estimated user's positions as they move.

Telephonic cells often overlap, and the numerous pings sent by a mobile phone cause signals to frequently bounce between cells, leading to noisy data. To address this, the available data are aggregated minute-wise, and they show only the position of the cell to which the phone was mostly connected every 60 seconds. Extensive preprocessing is then performed to distinguish cell bounces caused by genuine user movements from those resulting from intrinsic connection behavior.

Once the final trajectories are obtained, trajectory-specific covariates are retrieved, useful for the clustering procedure and the group interpretation process; see \cref{subfig:movement}. These include the origin-destination distance (OD; the distance from the origin to the destination in a beeline) and the departure time, used in the application of \cref{sec:application}, but other information, such as the total distance traveled, the average speed or the radius of gyration, could be useful.

The compositional trajectories are obtained by introducing the road-type data into the framework. 
If we had the exact positions, for example as GPS measurements, we could analyze a user's path in terms of the type of roads they travel, knowing exactly whether they stand on a sidewalk, on a cycle lane, or on a road, and from that we could draw conclusions on their behavior. However, since we employ the centers of mass of the telephonic cell coverage areas, we have to take into account the uncertainty of the approximation of the position. So, we simultaneously consider all the possible types of road that the user could be traveling on, using a compositional representation.
The simplex is the space of the proportions of the types of roads in the surroundings of the positions of the users. We count the meters of walkable pathways, cycle lanes and different types of roads in a circle of a radius of 200 meters around the estimated position for all measured locations obtained after preprocessing, as in \cref{subfig:roads}. The specific value for the radius has been chosen according to the size of the cell coverage areas.
The counts are normalized to obtain their proportion with respect to the total, that is, a point in the simplex. As a user moves, the proportion of roads in the surroundings changes, so a trajectory in the simplex is drawn; see \cref{subfig:comp_path}.

\begin{figure}[tp]
\centering
\refstepcounter{figure}
\addtocounter{figure}{-1}
    \begin{tikzpicture}
            \node[] at (-4, 4) (movement) {\includegraphics[trim = 13cm 6cm 12cm 6cm, clip, width = 0.45\linewidth]{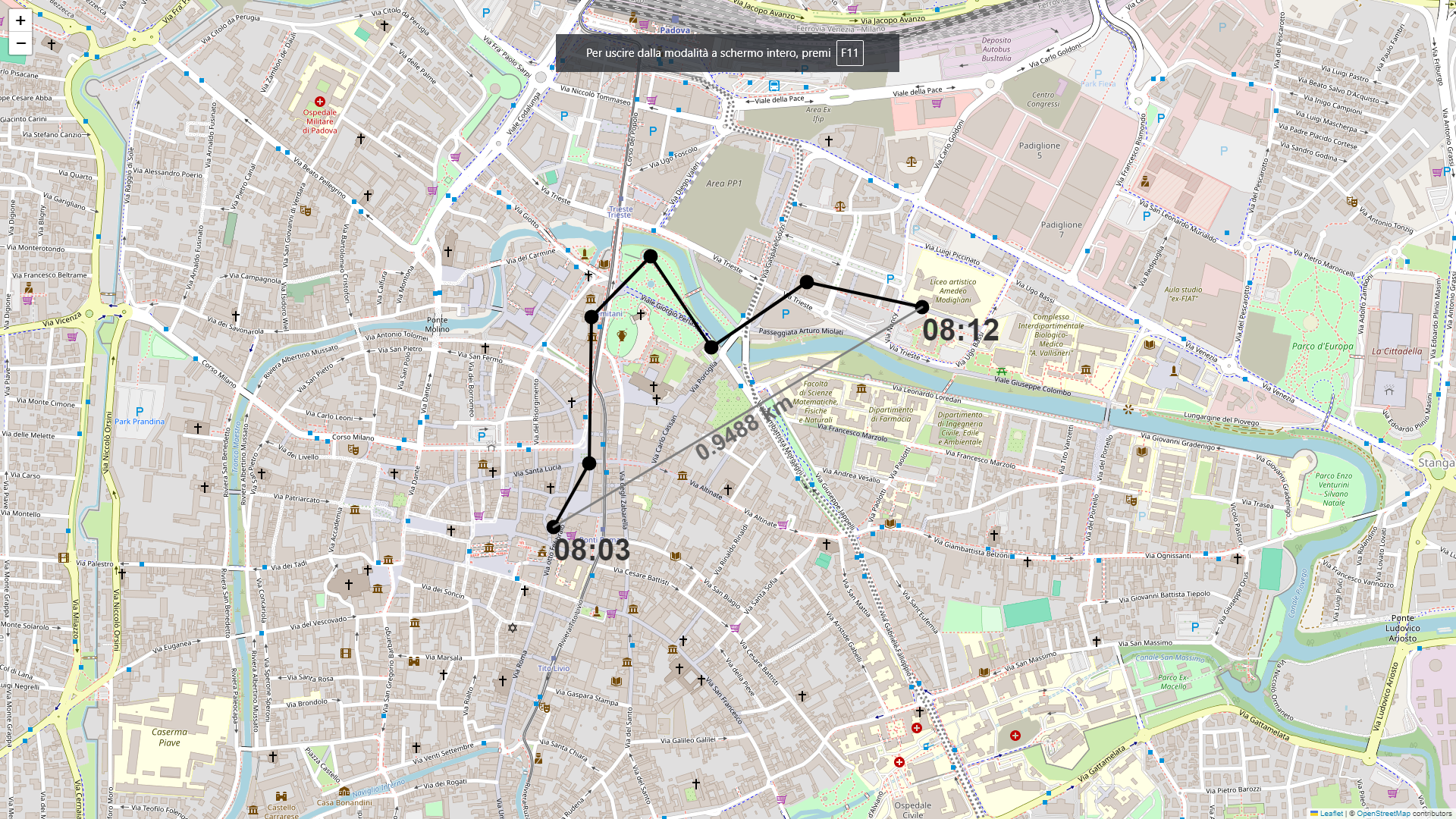}};
            
            \node[below=1ex of movement.south,inner sep=0pt] (captionA)
            {\begin{subfigure}{0.45\linewidth}
            \caption{Telephonic trajectory.}
            \label{subfig:movement}
            \end{subfigure}};
            
            \node[] at (5, 4) (map) {\includegraphics[trim = 13cm 3cm 13cm 3cm, clip, width = 0.35\linewidth]{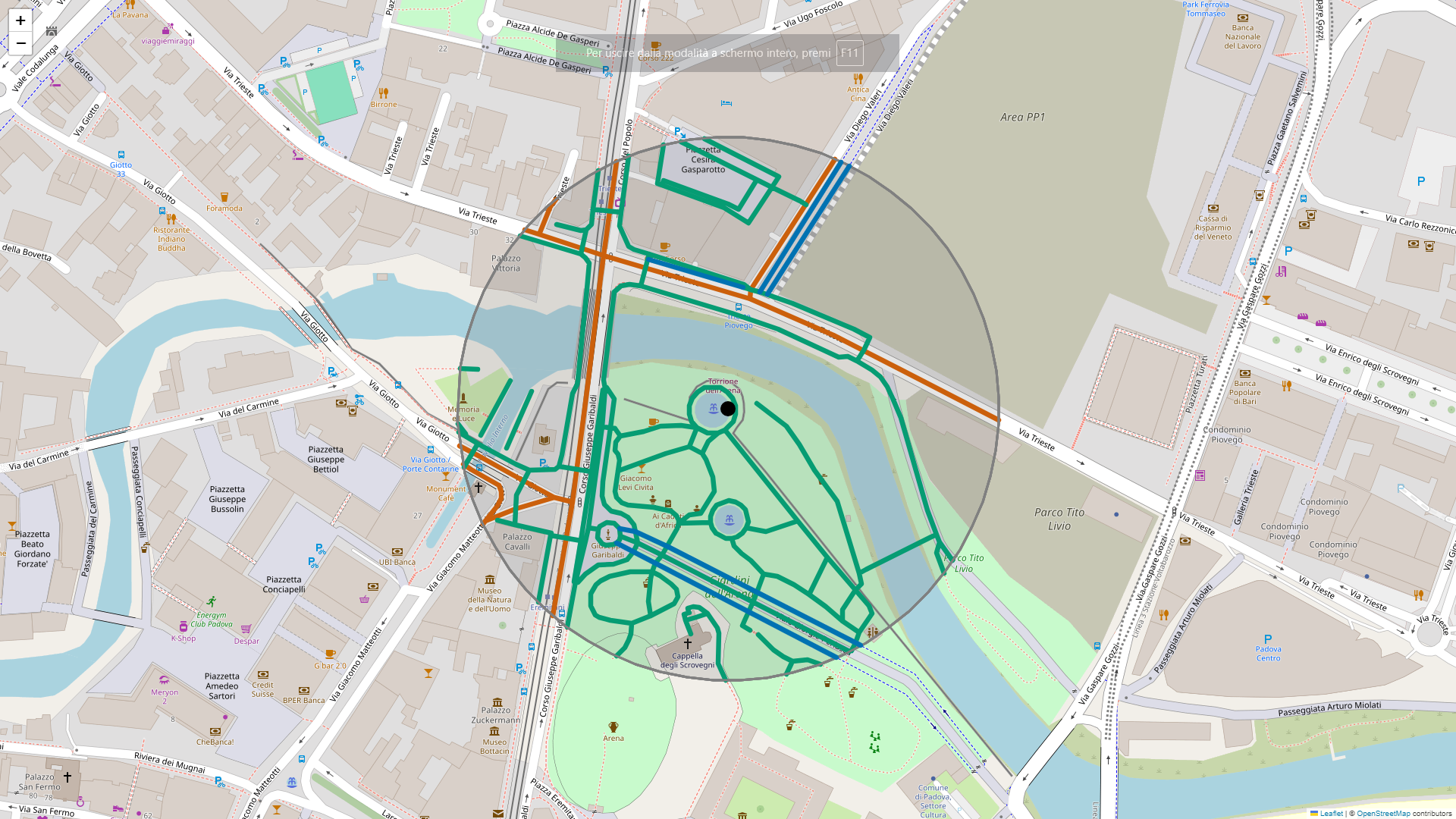}};
            \node[below= 1ex of map.south,inner sep=0pt] (captionB)
            {\begin{subfigure}{0.35\linewidth}
            \caption{Road representation.}
            \label{subfig:roads}
            \end{subfigure}};

            \node[below = 5ex of captionB.south] (table){
            \boxed{\scriptsize\begin{tabular}{cccc}
                 {\color{ViridisOrange}\textbf{Urban road}} & {\color{ViridisGreen}\textbf{Pedestrian}} & {\color{ViridisBlue}\textbf{Cycle lanes}} & \textbf{High-speed road} \\
                 \midrule
                 760.68m & 3250.50m & 521.36m & 0m \\
                 16.8\% & 71.6\% & 11.6\% & 0\% \\
            \end{tabular}}};
            
            \path [-stealth, bend left, line width = 2pt] (map.east) edge (table);
            
            \node[below = 10ex of table.west]  (symplex) {\includegraphics[trim = 1.4cm 1.9cm 1.8cm 2.8cm, clip, width = 0.45\linewidth]{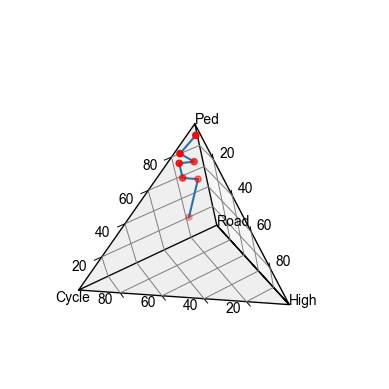}};
            \node[below= 1ex of symplex.south,inner sep=0pt] {
            \begin{subfigure}{0.35\linewidth}
            \centering
            \caption{Compositional trajectory.}
            \label{subfig:comp_path}
            \end{subfigure}
            };
            
            \node[draw, shape = circle, minimum size = 0.1cm, line width = 1pt, dashed] (circle) at (-5.05, 5.7){};
            \path [-stealth, bend left, line width = 1pt, dashed] (circle) edge (4.95, 4.05);
            
            \path [-stealth, bend right, line width = 2pt] (captionA.south) edge (-0.6, -4);
            
            \path [-stealth, bend left, line width = 2pt] (table.south) edge (1.5, -4.25);
            
    \end{tikzpicture}
    \caption{Compositional representation of a mock telephonic trajectory.}
    \label{fig:comp_repr}
\end{figure}
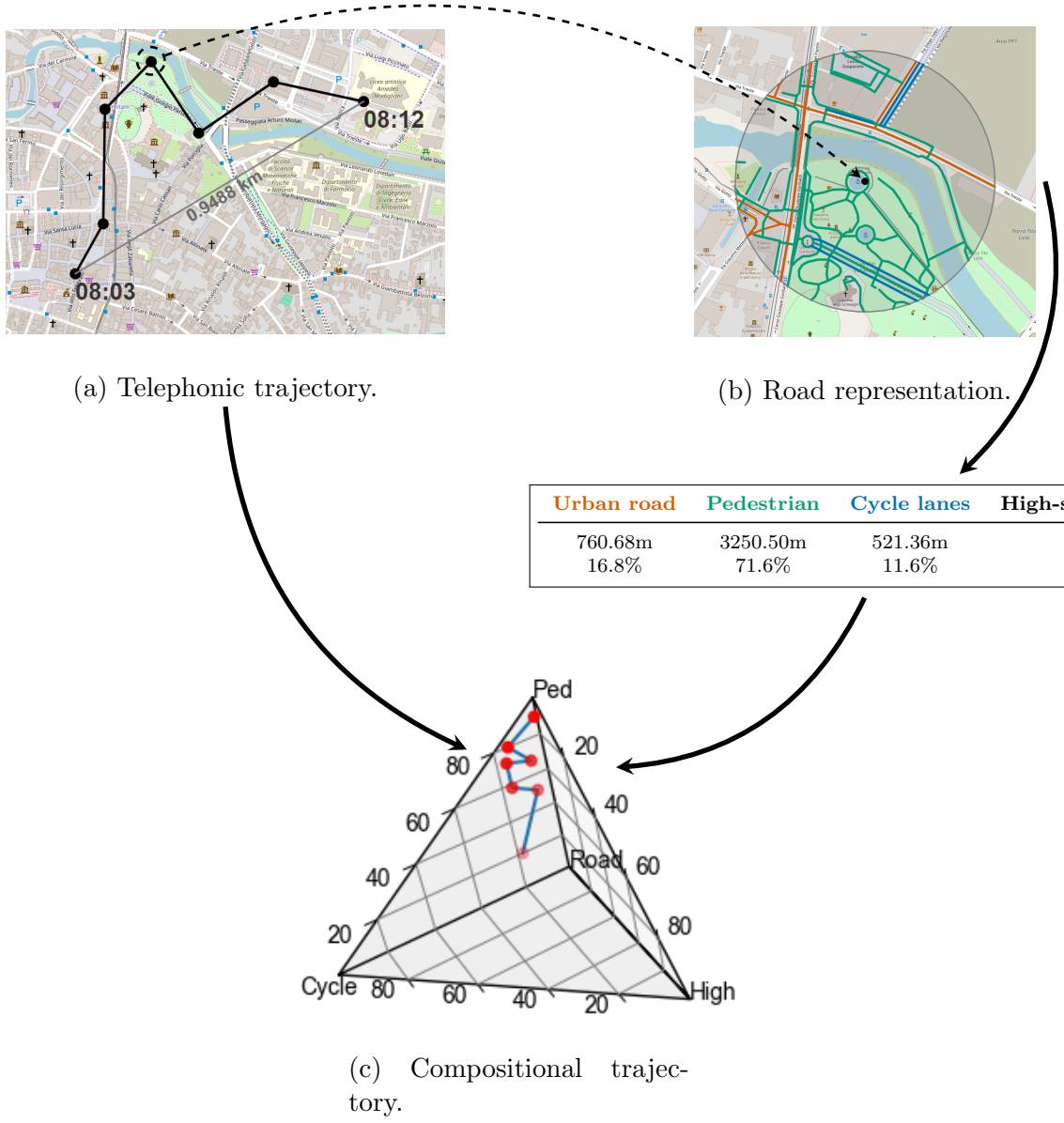

The measurements were collected in the province of Padova, Italy, from a sample of 399 individuals during a week in October 2022. 
Road-type data were collected from Geofabrik GmbH (\url{www.geofabrik.de}), which provides extracts from the OpenStreetMap project. 
The preprocessing and extraction of compositional trajectories have been implemented in Python \citep{rossum2009python}, while an R package \citep{R2025,Rcpp2025} with the code to perform the proposed model-based clustering procedure for compositional time series is available at \url{https://github.com/AndreaPanarotto/compStateSpace}.
Generative AI tools (GPT-4o, GPT-4.5, GPT-5) were used in the writing of this article in the form of language improvement tools and coding assistants to produce clearer figures. 

\section{Model definition} \label{sec:model_definition}

In the literature on compositional analysis, the proportions that are part of a composition, or total, are usually called `components'. Since we will also be dealing with components in the sense of elements that, interacting, construct a mixture of distributions, to avoid confusion, we will only address the compositional elements as `parts' or `proportions'.
Let $\mathbb{S}^{D-1} = \left\{\mathbf{y} \in \mathbb{R}_{>0}^D \mid \sum_{i=1}^D y_i = 1\right\}$ be the $D$-dimensional open simplex. Since the milestone work of \cite{aitchison1982statistical}, the usual way to deal with compositional data is to exploit logratio transformations that map the data from the constrained simplex space $\mathbb{S}^{D-1}$ to the unconstrained Euclidean space $\mathbb{R}^{D-1}$. Assuming a normal distribution on the transformed data allows one to induce flexible variance structures on the data, which is not possible when modeling data via common simplex-supported distributions, such as the Dirichlet distribution.

We adopt the isometric logratio (ilr) transformation \citep{egozcue2003isometric}. Given a composition $\mathbf{y} \in \mathbb{S}^{D-1}$, the corresponding ilr-transformed vector $\mathbf{z}={\rm ilr}(\mathbf{y}) = (z_1,\ldots,z_{D-1})$ is given by
\begin{eqnarray*}
    z_i = \left(\sqrt{\frac{i}{i+1}} \log \frac{\sqrt[i]{\prod_{j=1}^i y_j}}{y_{i+1}}\right),\qquad i=1,\ldots,D-1\;.
\end{eqnarray*} 
The term `isometric' comes from the property of preserving the distances between the Hilbert structure on the simplex and the Euclidean norm in $\mathbb{R}^{D-1}$ \citep{aitchison1986statistical, billheimer2001statistical}. This property is particularly appealing in a clustering framework, since preserving the distance structure avoids inducing further bias in the relationships between the observations after the transformation.

A compositional trajectory of length $n$ is an ordered sequence of compositional vectors, represented as $\mathbf{Y} = \left\{\mathbf{y}_t\right\}_{t=1,\ldots,n}$, with $\mathbf{y}_t\in\mathbb{S}^{D-1}$ for $t=1,\ldots,n\,$. We indicate with $N$ the total number of compositional trajectories, $\mathbf{Y}^{(1)},\ldots,\mathbf{Y}^{(N)}$, with respective lengths $n_1,\ldots,n_N$ not necessarily equal. To provide a modeling framework that also accounts for the noisy nature of the data, we employ a linear Gaussian state-space model \citep[e.g.][]{durbin2012time}, working with the ilr-transformed trajectories, that is, $\mathbf{Z}^{(i)} = \{\mathbf{z}_t^{(i)}\}_{t=1,\ldots,n_i}$ for $i = 1,\ldots,N$, where $\mathbf{z}_t^{(i)} = {\rm ilr}(\mathbf{y}^{(i)}_t)$ for $t=1,\ldots,n_i$. Zeros in an observed component (such as high-speed roads in \cref{fig:comp_repr}) are replaced by a small positive quantity to allow the logratio transformation to be well-defined. 

\subsection{Mixture of linear Gaussian state-space models}
\label{sec:mixture_state_space}

We model the trajectories through a mixture of linear Gaussian state-space models. We assume that each series belongs to a component $k=1,\ldots, K$, with $K$ being the total number of components. We indicate with $C_i$ the index random variable such that $C_i = k$ if the series $i$ belongs to component $k$. We assume that series in the same component share the same set of system matrices $\boldsymbol{\Theta}^{(k)} = \left(\TT^{(k)}, \QQ^{(k)}, \HH^{(k)}\right)$, so that, given an initial state $\boldsymbol{\alpha}^{(i)}_0$,
\begin{align*}
        \boldsymbol{\alpha}^{(i)}_t  &= \TT^{(k)} \boldsymbol{\alpha}^{(i)}_{t-1} + \boldsymbol{\xi}^{(i)}_t\;, & \boldsymbol{\xi}^{(i)}_t \sim \mathcal{N}\left(0,\  \QQ^{(k)}\right)\,,\\
        \mathbf{z}^{(i)}_t  &= \boldsymbol{\alpha}^{(i)}_t + \boldsymbol{\zeta}^{(i)}_t\;, &\boldsymbol{\zeta}^{(i)}_t\sim \mathcal{N}\left(0,\ \HH^{(k)}\right)\,,
\end{align*}
for all series $i$ with $C_i=k$. The observation-level errors $\boldsymbol{\zeta}_t^{(i)}$ represent the localization error due to telephonic measurement, while the state equations in $\boldsymbol{\alpha}_t^{(i)}$ describe the effective development of the trajectory. The component-specific system matrices are the elements on which the model-based clustering is performed. In contrast with a distance-based clustering, this approach allows us to cluster together series that could be distant in the simplex but that develop in time in a similar way in terms of the dependence between consecutive observations. 

We make some simplifying assumptions about the system matrices $\TT, \QQ$ and $\HH$. First, they are invariant over time.
In the case study, we deal with a large number of short series whose evolution in time is difficult to model. Moreover, this is a simple manner to account for the difference in length of the series, so that short series can borrow information from the longer series in the same cluster.
The second assumption is that the system matrices are diagonal, thus 
avoiding an over-parametrization of the model. 
In fact, the diagonal matrices act on the ilr-transformed variables, capturing well enough the evolution of the compositional series when they are transformed back in the simplex.

To assign each series to the different components, we assume a simple mixture of experts (ME) model \citep{gormley2019mixture}, which allows us to model the probability that a time series belongs to each component as a function of both its component-specific state-space likelihood and possible additional covariates. Let $k=1,\ldots, K$ indicate the components of the mixture and $\mathbf{x}_i$ be a vector of covariates of dimensions $P$ relative to series $i$. The likelihood function under the simple mixture of experts model is
\begin{equation}
    L\left(\mathbf{\Gamma}, \mathbf{\Theta};\, K\right)=p\left(\mathbf{Z}^{(1)},\ldots,\mathbf{Z}^{(N)};\, \mathbf{X}, \mathbf{\Gamma}, \mathbf{\Theta}\right)=\prod_{i=1}^N \sum_{k=1}^K \pi_k\left(\mathbf{x}_i;\, {\mathbf{\Gamma}}\right) L_S\left(\mathbf{\Theta}^{(k)};\mathbf{Z}^{(i)}\right).
    \label{eq:mixture_lik}
\end{equation}
Here, the $\pi_{k}$, as $k$ varies, indicate the component weights of the simple ME model, depending on the series covariates $\mathbf{x}_i$ and the current associated parameters in $\mathbf{\Gamma} = (\boldsymbol{\gamma}_1, \ldots, \boldsymbol{\gamma}_K)$, as 
\begin{equation}
    \pi_k\left(\mathbf{x}_i;\, {\mathbf{\Gamma}}\right)=\frac{\exp \left(\gamma_{k 0} + \mathbf{x}_i^\top\boldsymbol{\gamma}_{k 1}\right)}{\sum_{k^{\prime}=1}^K \exp \left(\gamma_{k^\prime 0} + \mathbf{x}_i^\top\boldsymbol{\gamma}_{k^{\prime} 1 }\right)}\,,
    \label{eq:comp_weight_params}
\end{equation}
where the vectors $\boldsymbol{\gamma}_k = ({\gamma}_{k0}, \boldsymbol{\gamma}_{k1}^\top)^\top,\ k=1,\ldots,K$, are of dimension $P+1$.
The component weights act as marginal probabilities of belonging to the corresponding component, using only the information of the covariates and not the series itself. 
Meanwhile, $L_S\left(\mathbf{\Theta}^{(k)};\mathbf{Z}^{(i)}\right)$ is the likelihood of the series $\mathbf{Z}^{(i)}$, computed via Kalman filter, given the current set of system matrices of the $k$-th component, $\mathbf{\Theta}^{(k)} = \{\TT^{(k)}, \QQ^{(k)},$ $ \HH^{(k)}\}$. 

\subsection{Parameter estimation}
\label{sec:KFS_est_Shumway}

The product-sum likelihood in Equation (\ref{eq:mixture_lik}) is difficult to maximize directly. Data augmentation is used to associate each series $i$ with a latent group membership indicator $\mathbf{w}_i = (w_{i1}, \ldots, w_{iK})^{\top}$, a vector of variables such that $w_{ik} = \mathbbm{1}(C_i = k)$, that is, $w_{ik}$ equals 1 if $\mathbf{Z}^{(i)}$ belongs to the $k$-th component and 0 otherwise. The complete-data likelihood is 
\begin{equation}
    \begin{aligned}
        L_c\left(\boldsymbol{\Gamma}, \boldsymbol{\Theta}, \mathbf{W} ; \,K\right)& = p\left( \mathbf{Z}^{(1)},\ldots,\mathbf{Z}^{(N)}, \mathbf{W};\, \mathbf{X}, \boldsymbol{\Gamma}, \boldsymbol{\Theta}\right)\\
        &=\prod_{i=1}^N \prod_{k=1}^K \left\{ \pi_k \left(\mathbf{x}_i;\, {\boldsymbol{\Gamma}}\right) L_S\left( \boldsymbol{\Theta}^{(k)};\, \mathbf{Z}^{(i)}\right) \right\}^{w_{ik}},
    \end{aligned}
    \label{eq:complete_lik}
\end{equation}
where the matrix $\mathbf{W}$ collects the vectors $\mathbf{w}_i\,,\ i=1,\ldots,N$.

An Expectation-Maximization algorithm \citep{dempster1977maximum} is adopted to iteratively optimize the likelihood. Initial estimates are given for the elements in $\mathbf{\Theta}^{(k)},\ k =1, \ldots, K$, and for $\mathbf{\Gamma}$ in the form of $(\boldsymbol{0}, \boldsymbol{\gamma}_2, \ldots, \boldsymbol{\gamma}_K)$, where the component weight parameters for group 1 are set to zero for identifiability reasons. At each iteration, in the outer expectation step, we use the current values of the parameters to estimate the latent group membership indicators of the series $i$ as the probability
\begin{equation}
    \hat{w}_{i k}\propto{\pi_k\left(\mathbf{x}_i;\, {\mathbf{\Gamma}}\right) L_S\left(\mathbf{\Theta}^{(k)};\,\mathbf{Z}^{(i)}\right)},\quad k= 1,\ldots,K\,.
    \label{eq:mixture_latent_w_estimate}
\end{equation}
While $\pi_k$ is the marginal assignment probability, the result of the product with the state-space likelihood of the series is proportional to the conditional probability of inclusion in the component given the data.

In the maximization step, the expectation of the complete-data loglikelihood is computed as the $Q$-function 
\begin{equation}
    \begin{aligned}
        Q(\mathbf{\Gamma}, \boldsymbol{\Theta}^{(1)}, \ldots, \boldsymbol{\Theta}^{(K)}) & = \mathbb{E}_{\boldsymbol{\Theta}_{curr}^{(k)}}\left[ L_c\left(\boldsymbol{\Gamma}, \boldsymbol{\Theta}, \mathbf{W} ; \,K\right)\mid \mathbf{Z}^{(1)},\ldots,\mathbf{Z}^{(N)}\right] \\ 
        & =  \sum_{i=1}^N \sum_{k=1}^K \hat{w}_{i k}\left\{\log \pi_k\left(\mathbf{x}_i ;\, {\mathbf{\Gamma}}\right) + \mathbb{E}_{\boldsymbol{\Theta}_{curr}^{(k)}}\left[\log L_S\left(\boldsymbol{\Theta}^{(k)};\mathbf{Z}^{(i)} \right)\right]\right\},
    \end{aligned}
    \label{eq:mixture_Q_function}
\end{equation}
where the likelihood of the state-space model is approximated by its expected value given the current parameters $\boldsymbol{\Theta}_{curr}^{(k)}$, as in \cite{shumway1982approach, shumway2000time}. The authors provide an EM algorithm to estimate the system matrices when analyzing a single series. It exploits the Kalman filter for computing the likelihood and the Kalman smoother to produce the estimates of the states given the entire trajectory information; the smoothed states are used for the estimation of the system matrices. Supplementary material A shows how we generalize their approach to our case, where we have $N$ time series, each of which evolves according to $\mathbf{\Theta}^{(k)}$ with probability $\hat{w}_{ik},\ i = 1,\ldots,N$. We perform one iteration of this procedure to update $\mathbf{\Theta}^{(k)},\ k =1, \ldots, K$.  
The parameters in $\mathbf{\Gamma}$ are updated, independently of those in $\mathbf{\Theta}$, by numerical maximization.

Clusters may be initialized randomly or using another clustering technique. In \cref{sec:application}, we employ the \texttt{dtwclust} package \citep{sarda2019time}, which performs time series clustering according to the Dynamic Time Warping (DTW) distance \citep{sakoe1978dynamic}. Such a distance is particularly suitable for computing similarities in different time scales thanks to the computation of an optimal matching. 

After all algorithms have converged, the ICL-BIC criterion (e.g. \citealp[Chapter~6]{mclachlan_finite_2000}; \citealp[Chapter~7]{fruhwirth-schnatter_finite_2006}) is used for model selection to simultaneously decide the optimal number of mixture components and which trajectory covariates are useful. In particular, once the set of covariates is fixed, we run the algorithm with a varying number of components $K = 2,\ldots, K_\text{max}$, and the final number of clusters is
\begin{equation}
    \hat{K} = \mathop{\operatorname{argmin}}_{K\in \{2,\ldots,K_{\text{max}}\}} n_P\log n_Z - 2 \log  L_c\left(\hat{\boldsymbol{\Gamma}}, \hat{\boldsymbol{\Theta}}, \hat{\mathbf{W}} ; K\right)\,,
    \label{eq:ICL-BIC_penalization}
\end{equation}
where the parameter estimates have been obtained from the procedures in \cref{sec:KFS_est_Shumway} and the values for $\hat{\mathbf{W}}$ are obtained from \cref{eq:mixture_latent_w_estimate}. In our specific case, the number of parameters $n_P$ is the number of the estimated parameters in $\hat{\boldsymbol{\Gamma}}$ plus the number of estimated diagonal elements of the system matrices in $\hat{\boldsymbol{\Theta}}$, $n_P = (K-1)(P+1) + 3K(D-1)$,
and the total number of observations is $n_Z = (D-1) \sum_{i=1}^N n_i$. BIC \citep{schwarz1978estimating} and ICL alone \citep{bachir2019inferring} have also been considered as criteria, but ICL-BIC appears to be more conservative in the choice of the number of groups and acts directly on the estimated complete likelihood we use in the estimation process.

Supplementary Material B shows a simulation study to assess the validity of the clustering procedure.

\section{Urban mobility in Padova}
\label{sec:application}
\noindent
The method is applied to $N=3015$ trajectories in the municipality of Padova, Italy, and its immediate surroundings, gathered between 10 pm on Sunday, 9 October 2022 and 10 pm on Sunday, 16 October 2022. We keep trajectories with at least 5 observations (time points), which implies that they have a minimum duration of 5 minutes, and consider a simplex size $D = 4$ as we measure the following road types: fast-paced roads, such as highways or ring roads; other types of road, including all urban roads; cycle lanes; and walkable pathways. The length of the series varies between 5 and 47, with a median of 11.

\subsection{Clustering of urban trajectories}
Our first research objective is to investigate whether urban mobility trajectories can be interpreted as macro-behavioral patterns characterized by distinct uses of the transportation infrastructure. We separate the analysis of the 2301 weekday trajectories from that of the 714 weekend trajectories, since the mass mobility behavior is very different between the two categories of days. As external covariates, we consider the OD distance for all trajectories and an indicator of whether the departure occurred during rush hours (7 am to 9 am, 12:30 pm to 2 pm, and 4 pm to 8 pm).

The ICL-BIC criterion in \cref{eq:ICL-BIC_penalization} is used to select the final number of clusters. However, we do not fit the models independently of each other as we vary $K$. In fact, we observed that, with real data and low $K$, the algorithm might converge to local maxima of the likelihood, depending on the starting point. With larger $K$, the algorithm is more stable, so we exploited this stability. In particular, we proceed iteratively, starting from a large number of components $K_{\text{max}}$, and after convergence of the EM algorithm, we remove one component from the mixture until $K=2$. This allows us to provide a good initialization for cases with a low number of groups. This backward elimination procedure is described in Supplementary Material C.

\begin{figure}[htp]
\centering
    \begin{subfigure}{0.55\linewidth}
    \centering
    \includegraphics[width=\linewidth]{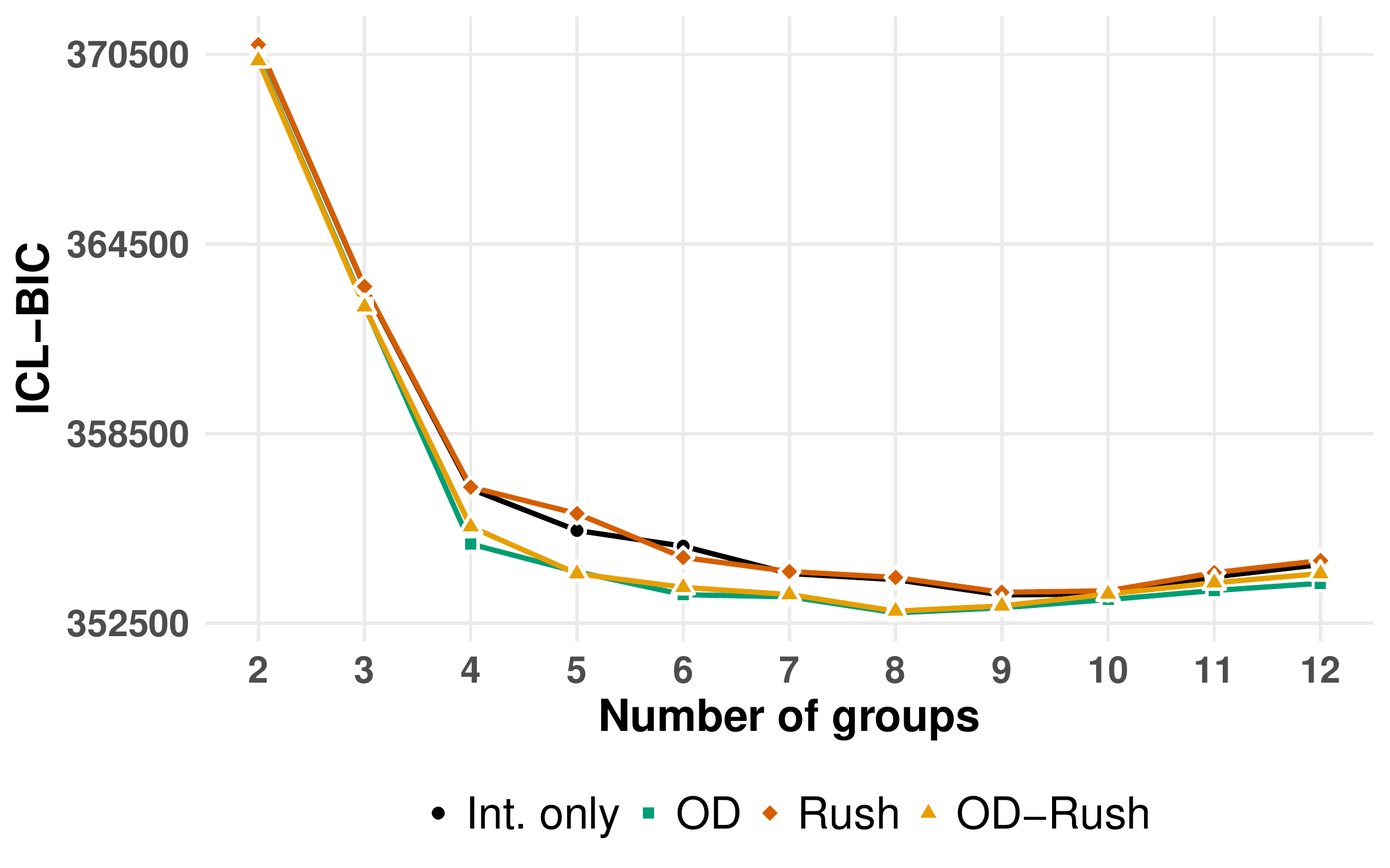}
    \caption{Model selection based on ICL-BIC.}
    \label{subfig:BicWD}
    \end{subfigure}
    \hfill
    \begin{subfigure}{0.4\linewidth}
    \centering
    \includegraphics[width=\linewidth]{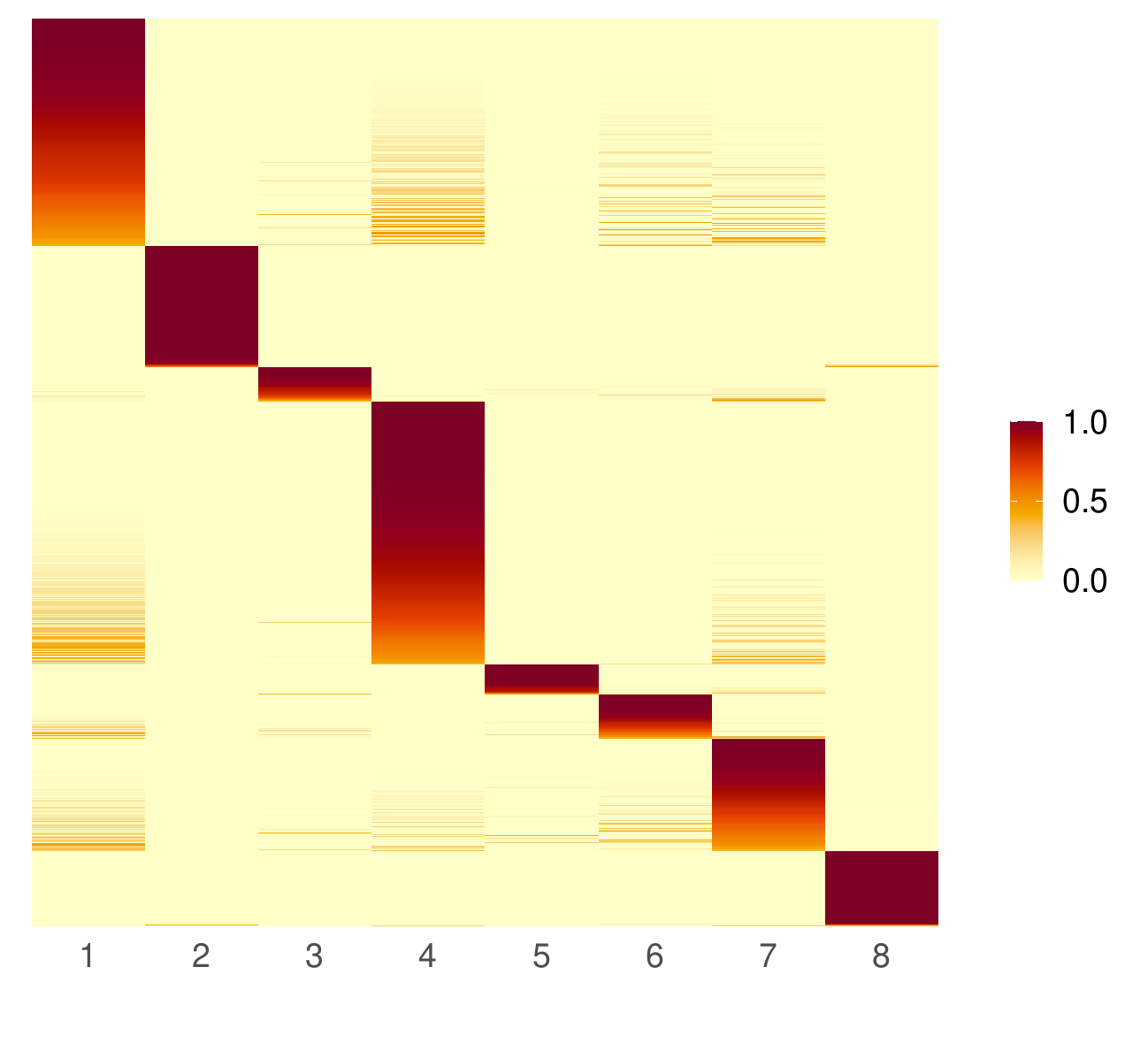}
    \caption{Assignment probabilities.}
    \label{subfig:assign_heatmap}
    \end{subfigure}
    
    \caption{Weekday analysis model selection and estimated assignment probabilities ($\hat{w}_{ik}$) to the components (on the columns) for the weekday series (on the rows).}
\end{figure}

We start from $K_\text{max}=12$. The results of the model selection for the weekdays are shown in \cref{subfig:BicWD}.

The optimal number of clusters in the weekday case is 8. The best model is the one with only the OD distance as a covariate. This characteristic was expected to be useful, as covering longer distances requires travel modes (cars or public transports) that are forced to move on urban or fast-paced roads, while shorter distances can be covered by walking or riding micromobility devices that have access to bike lanes. It is interesting that the rush hour information does not bring particular benefits to the analysis, meaning that the departure time or the traffic conditions do not have a very strong effect on the path selected by users.

Once the optimal model has been determined, we use its outcome to complete the clustering procedure. In particular, the latent group membership estimators act as posterior assignment probability to each component, so we can assign each series $i$ to the cluster having the largest value for such estimator,
\begin{equation*}
    \hat{C}_i = \mathop{\operatorname{argmax}}_{k=1,\ldots,K} \hat{w}_{ik}.
\end{equation*}

\cref{subfig:assign_heatmap} is a heatmap showing the assignment probability of the 2301 weekday series to the 8 clusters obtained by the optimal model. For clarity, the series have been ordered first according to the assigned cluster and then according to the value of the membership estimator for the corresponding component, as a measure of the strength of the probability to be assigned to that group. We observe that the series are well separated across the clusters, with the vast majority of them having an assignment probability to their component close to 1. The series with a lower maximum assignment probability are mostly divided between the largest clusters, which makes sense since those components have a higher prior assignment probability due to their size, but also because those clusters actually show similar characteristics, as we will see in the following.   

\begin{figure}[tp]
\centering

\begin{subfigure}{0.47\linewidth}
\centering
\includegraphics[width=\linewidth]{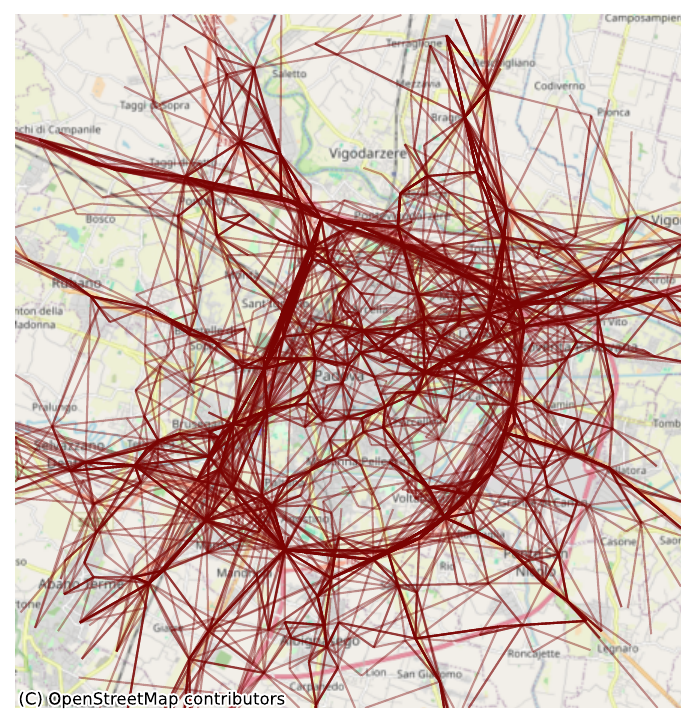}
\caption{Cluster 1, 578 trajectories, $\hat{\gamma}_{1,OD}=0$.}
\label{fig:map_cl1}
\end{subfigure}
\hfill
\begin{subfigure}{0.47\linewidth}
\centering
\includegraphics[width=\linewidth]{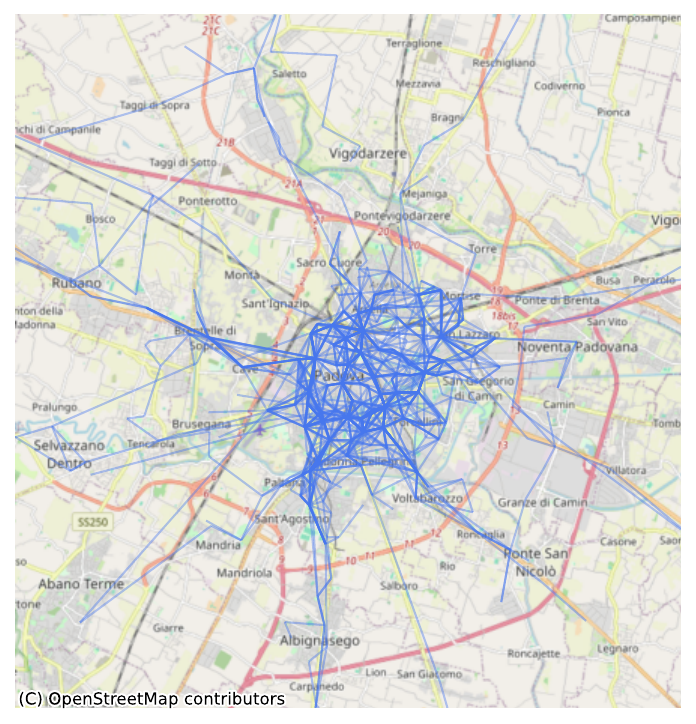}
\caption{Cluster 2, 306 trajectories, $\hat{\gamma}_{2,OD}=-3.51$.}
\label{fig:map_cl2}
\end{subfigure}

\medskip

\begin{subfigure}{0.47\linewidth}
\centering
\includegraphics[width=\linewidth]{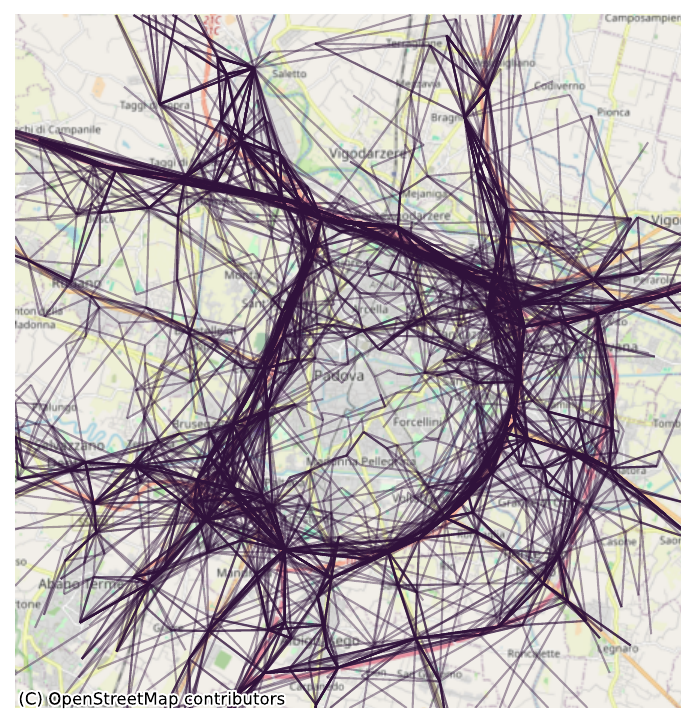}
\caption{Cluster 4, 666 trajectories, $\hat{\gamma}_{4,OD}=0.39$.}
\label{fig:map_cl4}
\end{subfigure}
\hfill
\begin{subfigure}{0.47\linewidth}
\centering
\includegraphics[width=\linewidth]{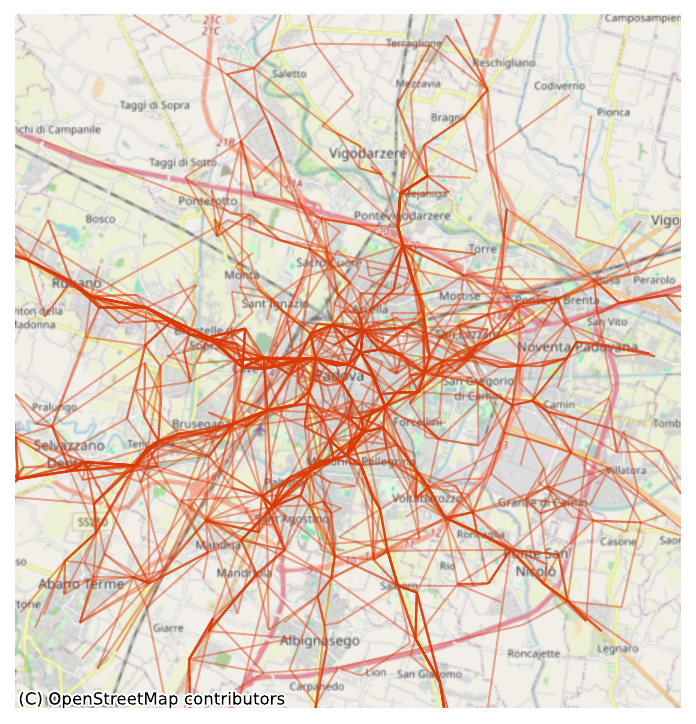}
\caption{Cluster 7, 284 trajectories, $\hat{\gamma}_{7,OD}=-0.67$.}
\label{fig:map_cl7}
\end{subfigure}
\caption{Trajectories plotted on map for a subset of clusters.}
\label{fig:map_cl}
\end{figure}

\cref{fig:map_cl} shows the superposition of the trajectories on the map for four of the eight clusters. Annotations are added on the cardinalities of the clusters and the estimate of the component weight parameter $\hat{\gamma}_{k,OD}$, relative to the origin-destination distance, and useful for interpretation. 
Cluster 2 is the one with the lowest value $\hat{\gamma}_{2,OD}$ and, in fact, collects shorter trajectories that are located entirely in the inner part of the city. These behaviors are typical of residents working in the city, but also of tourists covering short distances between points of interest. Cluster 4 is the largest in cardinality and the one with the highest estimated value $\hat{\gamma}_{4,OD}$, and contains longer trajectories on the exterior highway and the beltway. Many of the trajectories in this cluster belong to people who live and work outside of Padova, and exploit the fast paced roads for their commutes without interacting much with the city. Cluster 1, the second in both the cardinality of the group and the value of $\hat{\gamma}_{1,OD}$, is composed of trajectories moving between the ring road and the inner city, and less on the motorway. Cluster number 7 shows movements between the inner city and the periphery with the other close municipalities. These two clusters contain people who live in the periphery and need to commute long distances to work in the city, with a different exploitation of the ring road. 

\begin{figure}[tp]
\centering
\begin{subfigure}{0.47\linewidth}
\centering
\includegraphics[width=\linewidth]{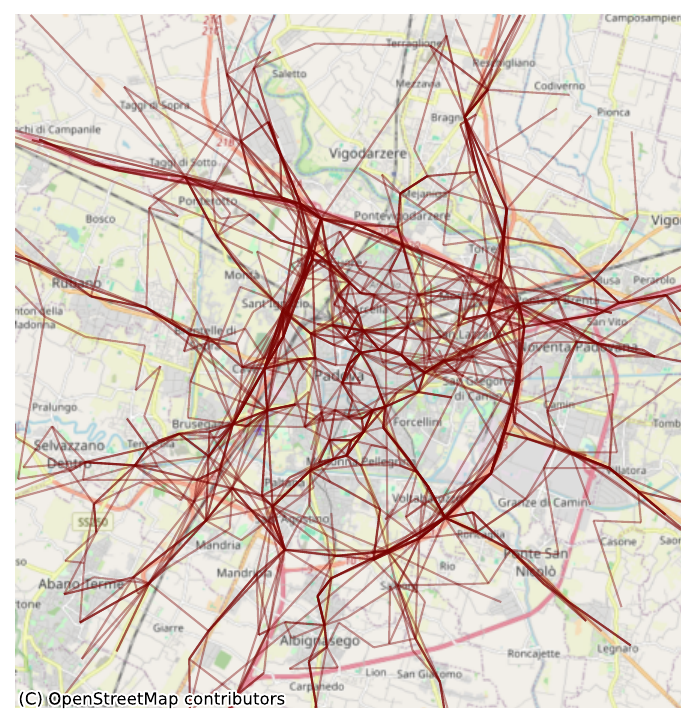}
\caption{Cluster 2, 190 trajectories, $\hat{\gamma}_{2,OD}=0$.}
\label{fig:map_cl_2WE}
\end{subfigure}
\hfill
\begin{subfigure}{0.47\linewidth}
\centering
\includegraphics[width=\linewidth]{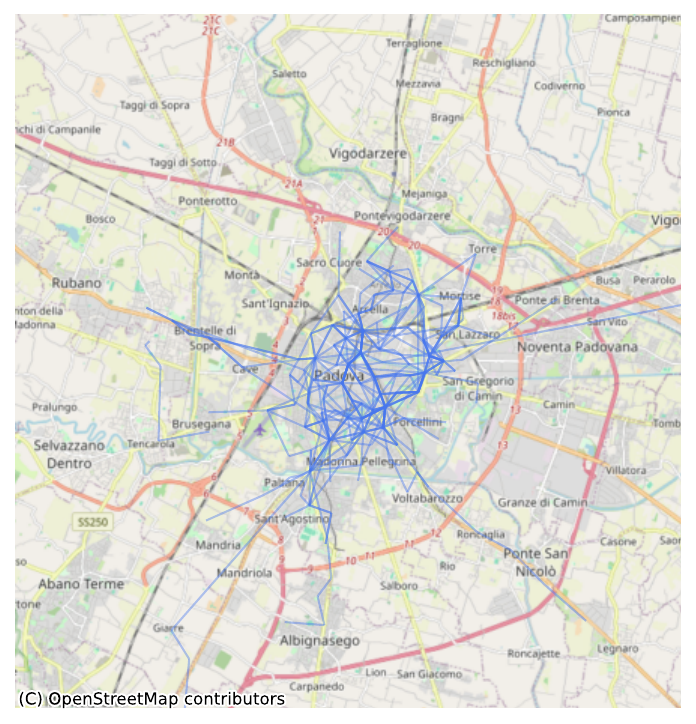}
\caption{Cluster 5, 84 trajectories, $\hat{\gamma}_{5,OD}=-3.73$.}
\label{fig:map_cl_5WE}
\end{subfigure}
\medskip

\begin{subfigure}{0.47\linewidth}
\centering
\includegraphics[width=\linewidth]{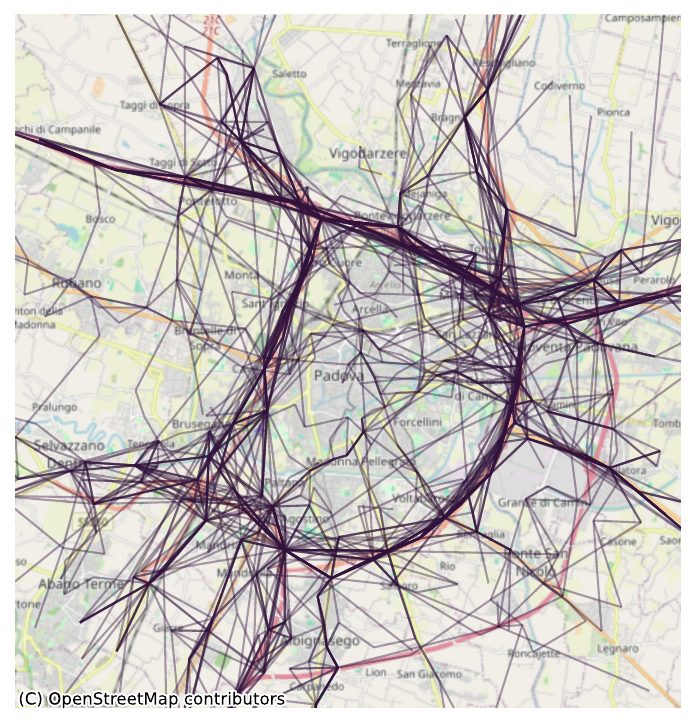}
\caption{Cluster 3, 244 trajectories, $\hat{\gamma}_{3,OD}=0.44$.}
\label{fig:map_cl_3WE}
\end{subfigure}
\hfill
\begin{subfigure}{0.47\linewidth}
\centering
\includegraphics[width=\linewidth]{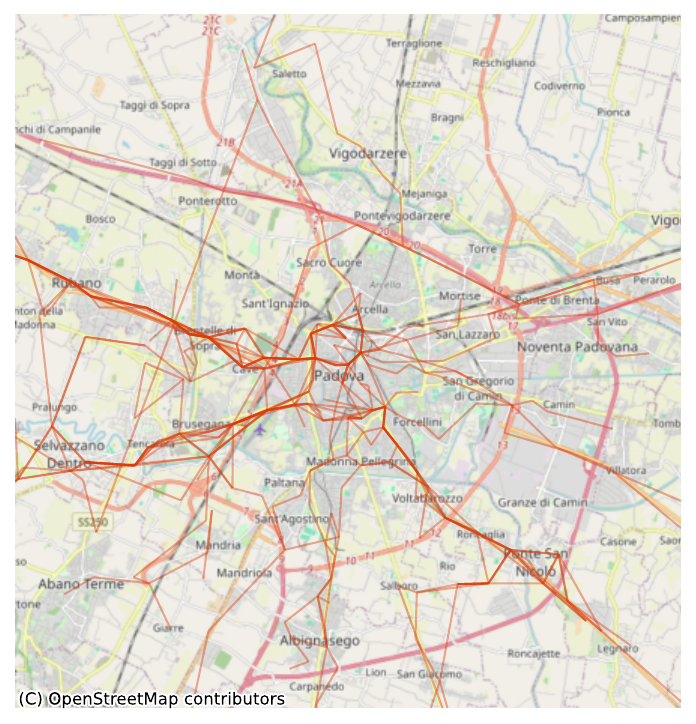}
\caption{Cluster 7, 53 trajectories, $\hat{\gamma}_{7,OD}=-0.23$.}
\label{fig:map_cl_7WE}
\end{subfigure}
\caption{Trajectory clusters plotted on map for the weekend analysis.}
\label{fig:map_cl_WE_14}
\end{figure}

In the analysis of the 714 weekend trajectories, information on rush hours was not included as there are no real rush hours during the weekend in terms of traffic jams. Again, the OD distance is selected as a covariate for the best model; see Supplementary Material D. The number of detected clusters is 7, less than the 8 in the weekday analysis.
This suggests a mobility behavior simpler to describe in the weekend, probably due to the lower number of total trajectories than to an effective different use of roads or travel modes. However, it is possible to delineate some similarities between the previously discussed weekday clusters and the weekend clusters in \cref{fig:map_cl_WE_14}, 
both visually and in terms of the corresponding number of trajectories and component weight parameters.

In particular, cluster 2 shows that, as for cluster 1 during the weekdays, around a quarter of the detected traffic consists of movements between high-speed roads and the inner part of the city; see 
Figures~\ref{fig:map_cl1}~and~\ref{fig:map_cl_WE_14}\subref{fig:map_cl_2WE}. We assign $\hat{\gamma}_{2,WE}=0$ and treat it as a baseline, as was done for $\hat{\gamma_1}$. This is legit since the component weight parameters are identified up to an additive constant.
Cluster 5 on the weekend is comparable to cluster 2 on weekdays, with movements in the inner part of the city and a small component weight parameter $\hat{\gamma}_{5,OD}^{WE}$; see 
Figures~\ref{fig:map_cl2}~and~\ref{fig:map_cl_WE_14}\subref{fig:map_cl_5WE}. 
Cluster 3 is similar to cluster 4 during the weekdays, but comparing  Figure~\ref{fig:map_cl_WE_14}\subref{fig:map_cl_3WE} with 
\cref{fig:map_cl4}, 
we observe a difference in the use of the south-eastern highway, A13. This highway connects Veneto and Emilia-Romagna, two highly active regions in terms of transport and trade, and is therefore busier on weekdays due to work-related traffic, particularly heavy vehicles such as trucks. However, when considering the proportion of trajectories within the respective datasets, the weekend cluster appears comparatively more relevant: the weekday cluster includes 666 trajectories out of 2301 total, whereas the analogous weekend cluster contains 244 trajectories out of 714 total. Although the A13 corridor is less prominent during weekends, the higher relative weight of the weekend cluster highlights the importance of the other major routes involved, namely the A4/E55 highway that runs east-west north of the city and the ring road, which remain important crossing points also outside of weekday commuting patterns. 
Cluster 7 in Figure~\ref{fig:map_cl_WE_14}\subref{fig:map_cl_7WE} is analogous to cluster 7 in 
\cref{fig:map_cl7},
which shows the trajectories between Padova and the periphery; during the weekend, this type of traffic is more concentrated in neighboring municipalities in the south and west.

Overall, the clustering procedure identifies a limited number of recurrent mobility behaviors characterized by distinct patterns of transportation infrastructure usage. Rather than analyzing thousands of individual trajectories separately, the proposed framework provides a compact representation of urban mobility demand, where each cluster corresponds to a macro-behavioral profile with a clear interpretation in terms of trip length, spatial organization and road usage. This representation can be exploited to study aggregated origin-destination structures, infrastructure utilization patterns and traffic loads associated with specific mobility behaviors, while preserving the main structural information contained in the original dataset. In this sense, the clusters should not be interpreted merely as statistical groups, but as data-driven mobility segments that provide an interpretable intermediate layer between individual trajectories and network-level mobility analyses.

The maps for the remaining clusters are in the Supplementary Material D.

\subsection{Intracluster analysis}

Our second research objective is to characterize the internal spatial organization of the detected macro-behavioral patterns through the distribution of origins and destinations throughout the urban area. This analysis provides additional insight into the functional interpretation of the clusters by highlighting recurrent departure and arrival areas at different times of the day. In the following, we focus on the weekday clusters displayed in \cref{fig:map_cl}.

\begin{figure}[tp]
\centering

\begin{subfigure}{0.47\linewidth}
\centering
\includegraphics[width=\linewidth]{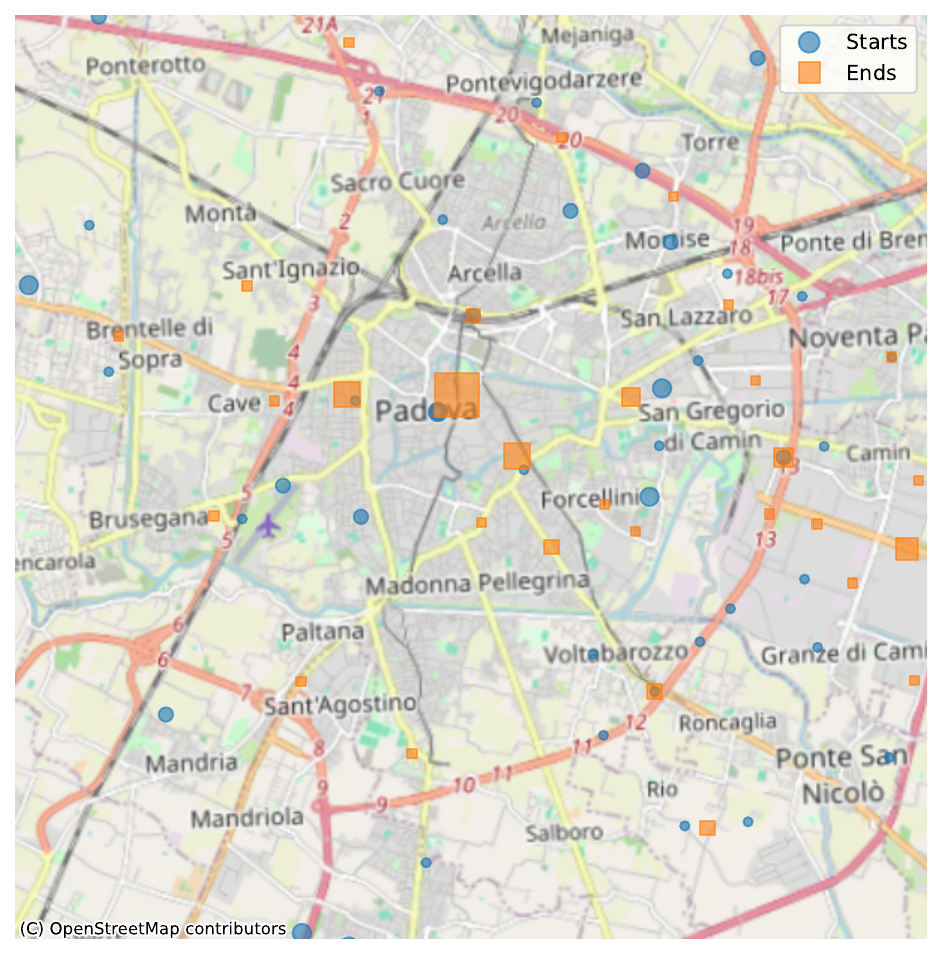}
\caption{Cluster 1 between 6:00 and 9:00.}
\label{fig:bubble_morning_1}
\end{subfigure}
\hfill
\begin{subfigure}{0.47\linewidth}
\centering
\includegraphics[width=\linewidth]{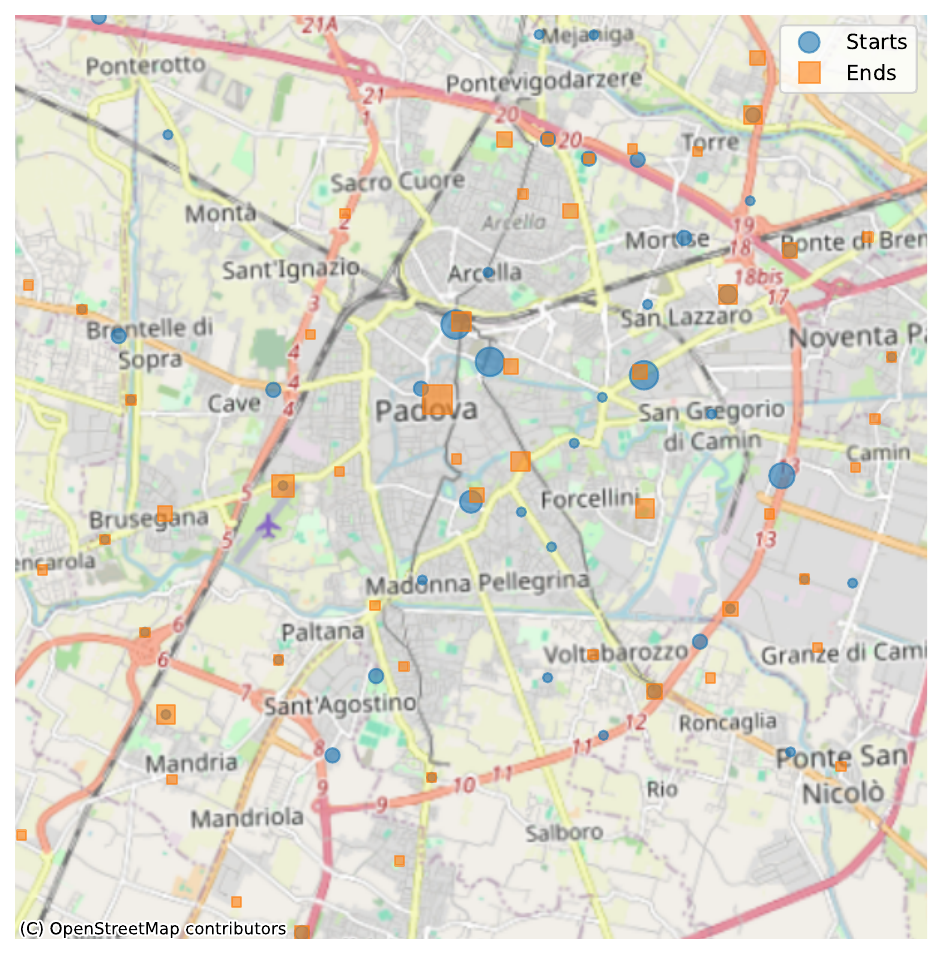}
\caption{Cluster 1 between 17:00 and 20:00.}
\label{fig:bubble_evening_1}
\end{subfigure}

\medskip

\begin{subfigure}{0.47\linewidth}
\centering
\includegraphics[width=\linewidth]{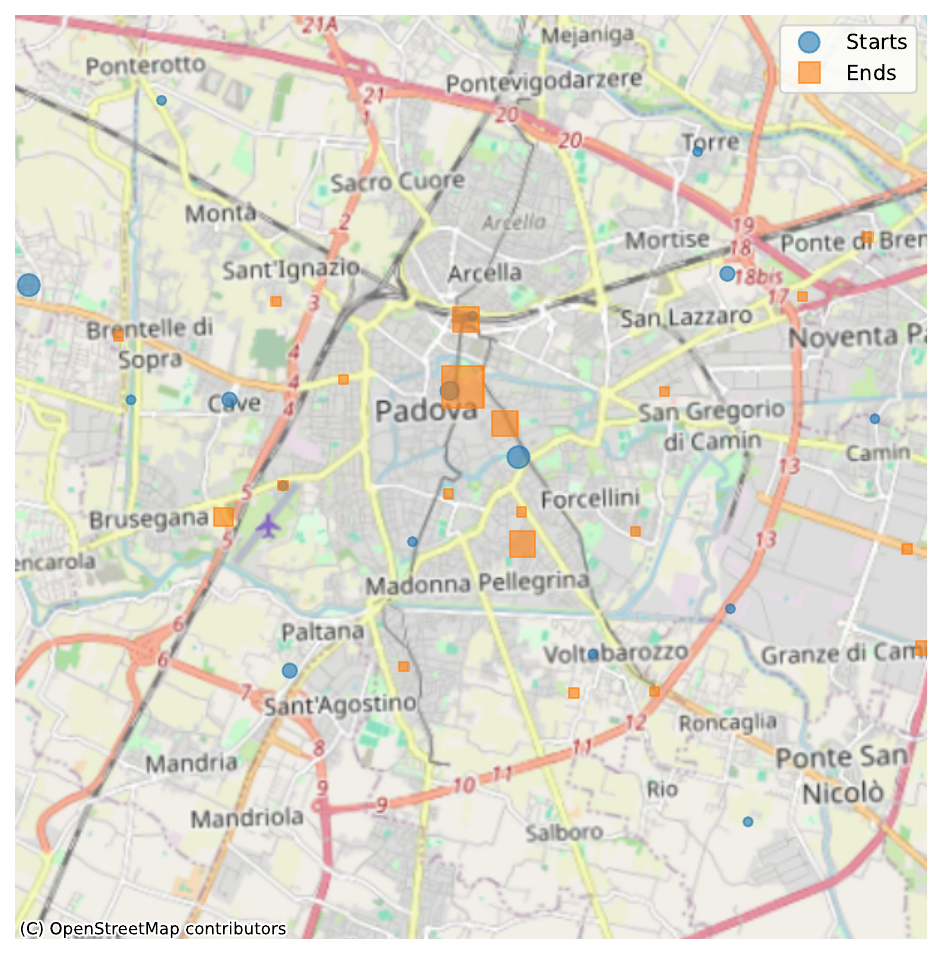}
\caption{Cluster 7 between 6:00 and 9:00.}
\label{fig:bubble_morning_7}
\end{subfigure}
\hfill
\begin{subfigure}{0.47\linewidth}
\centering
\includegraphics[width=\linewidth]{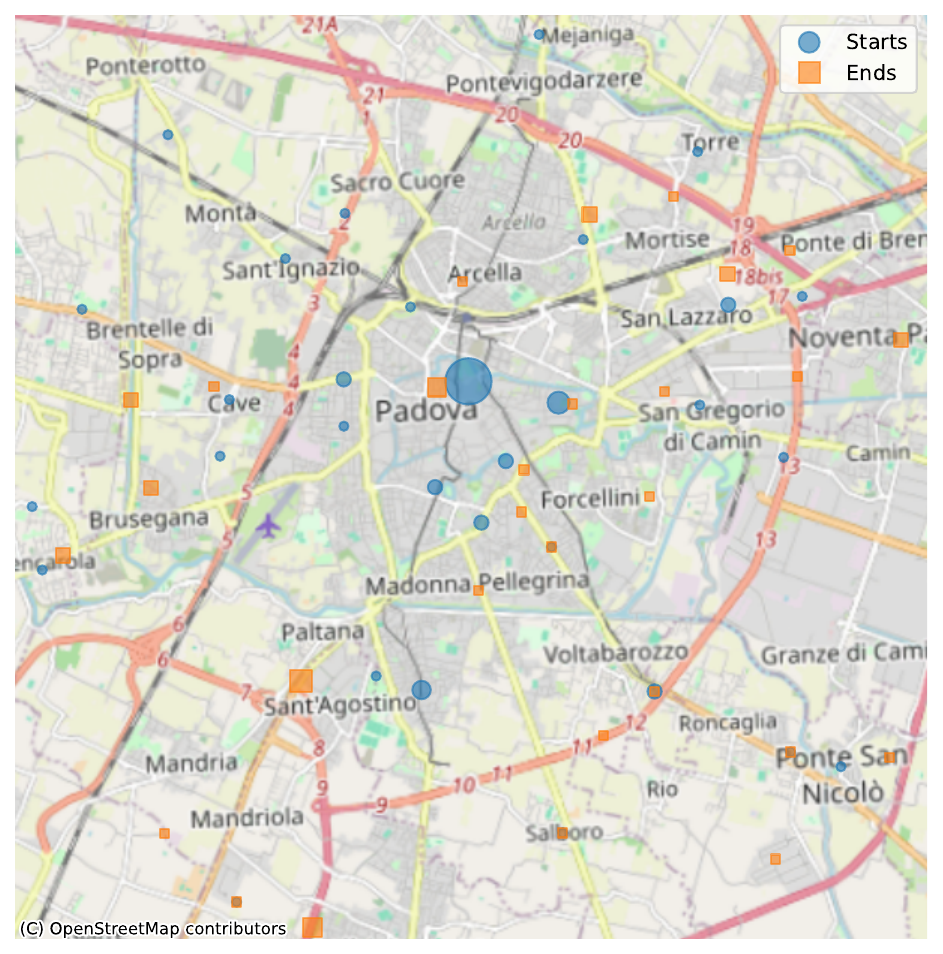}
\caption{Cluster 7 between 17:00 and 20:00.}
\label{fig:bubble_evening_7}
\end{subfigure}

\caption{Origin and destination concentration areas for clusters 1 and 7 during morning and evening periods.}
\label{fig:bubble_17}
\end{figure}

The bubbleplots in \cref{fig:bubble_17} summarize the main origin and destination areas for clusters 1 and 7, which were previously associated with commuting behaviors involving residents of the peripheral municipalities surrounding Padova. The temporal structure of the detected points confirms this interpretation. During the morning, the main departure locations are concentrated in the peripheral areas, while the main destinations are located in the city center. The opposite behavior is observed in the late afternoon and evening, when the trajectories predominantly move from the center to residential areas outside the city.

Cluster 1 also highlights the role of the industrial district located in the eastern part of Padova, which appears as an important destination during the morning and as a departure area in the evening. The cluster also contains users residing within the city who exploit the ring road and the highway network to leave the urban area in the morning and return in the evening, as suggested by the concentration of endpoints near the boundaries of the study domain. 

\begin{figure}[tp]
\centering

\begin{subfigure}{0.47\linewidth}
\centering
\includegraphics[width=\linewidth]{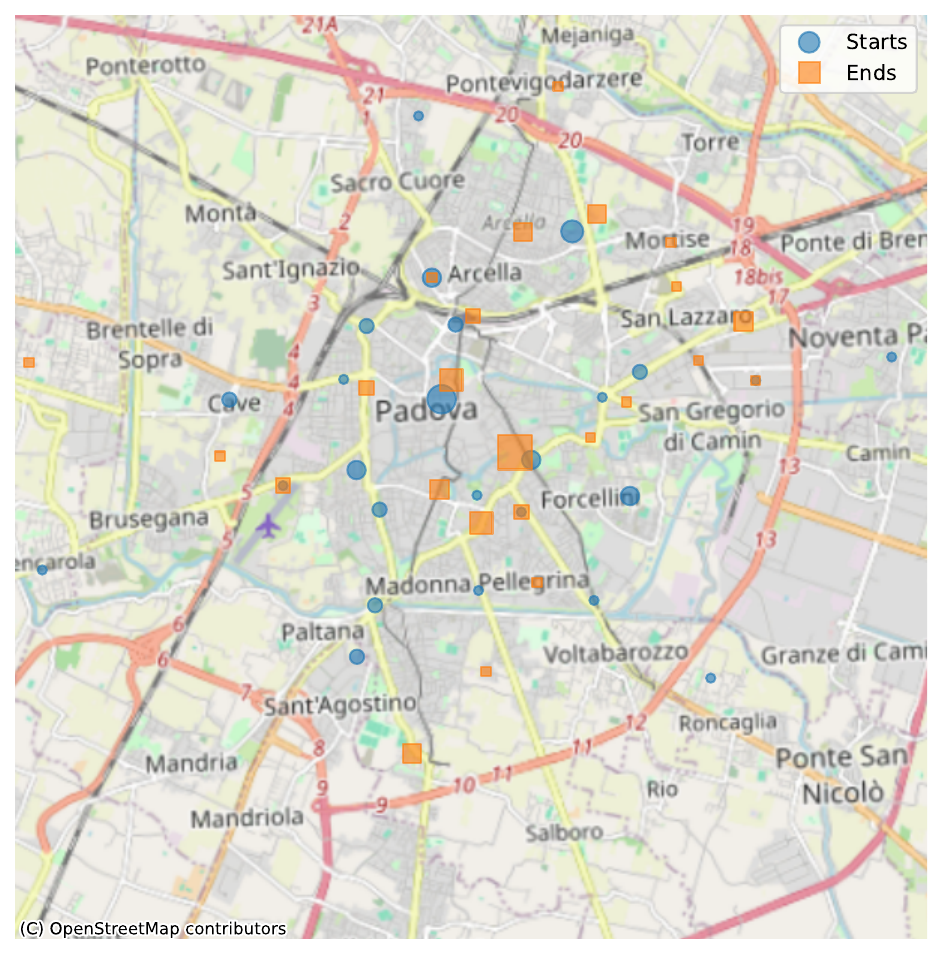}
\caption{Cluster 2 between 6:00 and 9:00.}
\label{fig:bubble_morning_2}
\end{subfigure}
\hfill
\begin{subfigure}{0.47\linewidth}
\centering
\includegraphics[width=\linewidth]{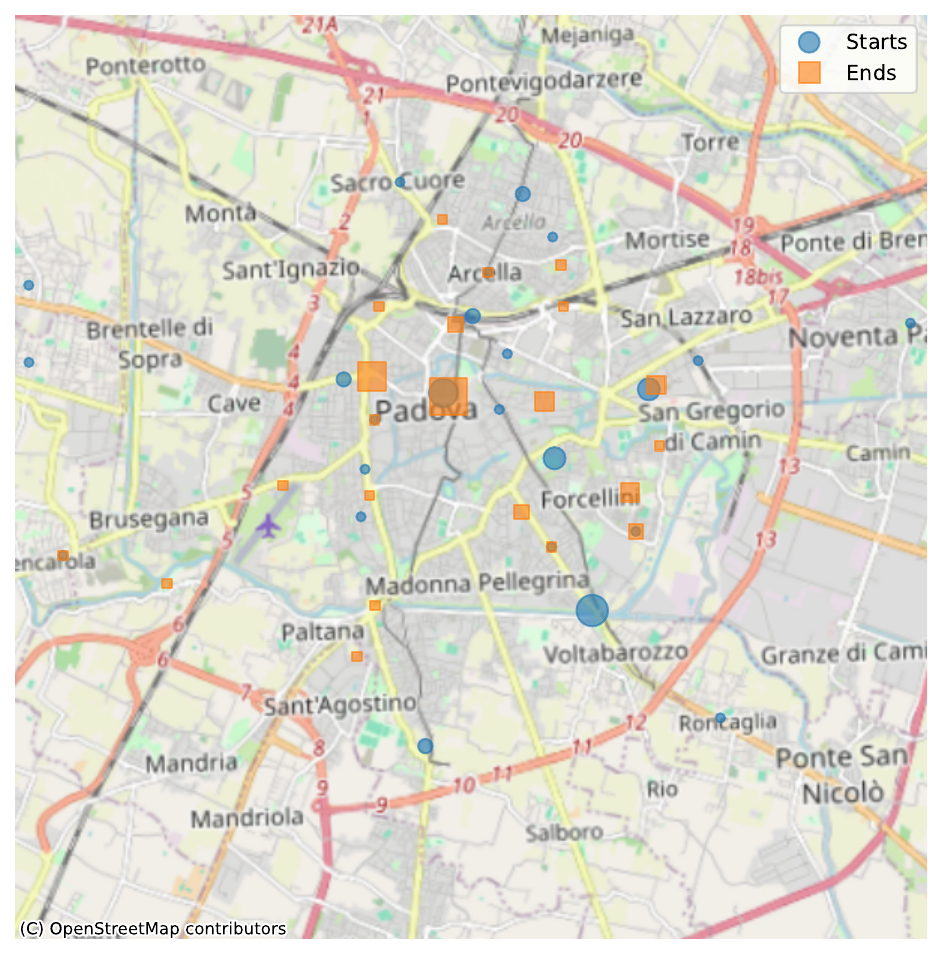}
\caption{Cluster 2 between 9:00 and 12:00.}
\label{fig:bubble_late_morning_2}
\end{subfigure}

\medskip

\begin{subfigure}{0.47\linewidth}
\centering
\includegraphics[width=\linewidth]{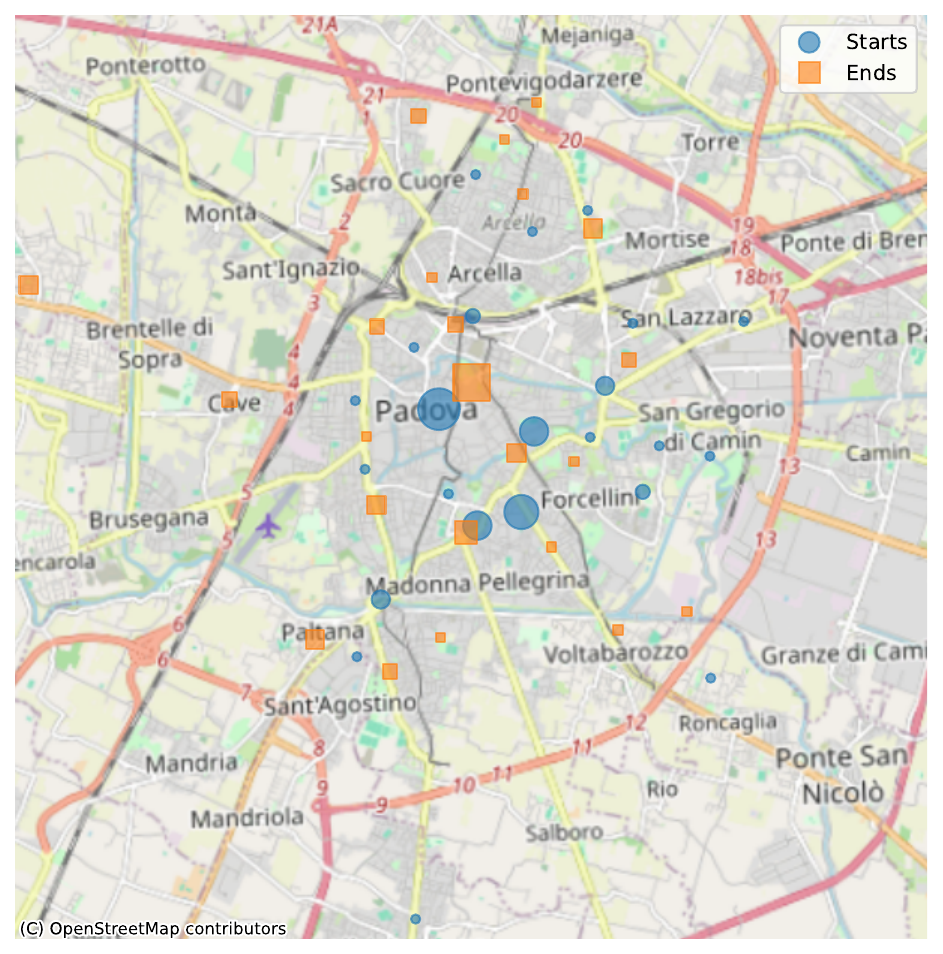}
\caption{Cluster 2 between 12:00 and 15:00.}
\label{fig:bubble_afternoon_2}
\end{subfigure}
\hfill
\begin{subfigure}{0.47\linewidth}
\centering
\includegraphics[width=\linewidth]{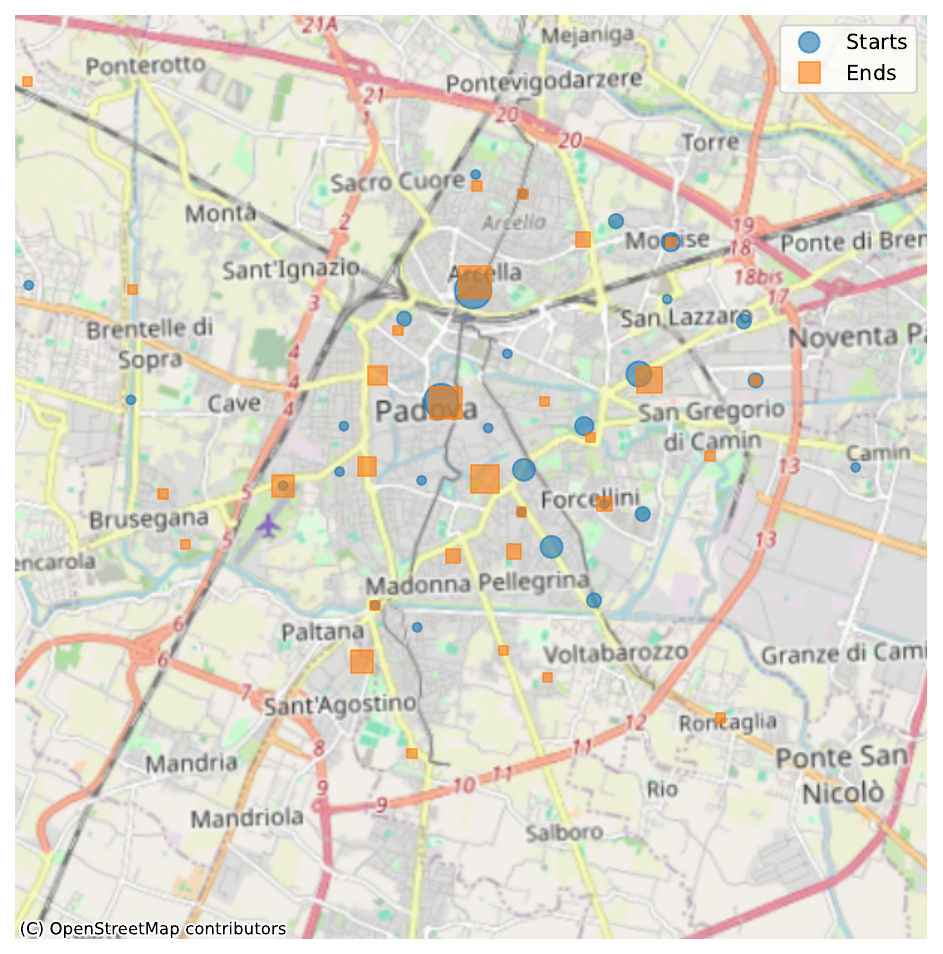}
\caption{Cluster 2 between 17:00 and 20:00.}
\label{fig:bubble_evening_2}
\end{subfigure}

\caption{Origin and destination concentration areas for cluster 2 across different times of the day.}
\label{fig:bubble_2}
\end{figure}

The intracluster structure of cluster 2, shown in \cref{fig:bubble_2}, is particularly relevant from the perspective of sustainable mobility and urban public transport. During the morning and evening commute periods, movements within the city center appear to be organized predominantly along a north-south axis, approximately corresponding to the current tram line and connecting major urban attractions such as the Saint Anthony Basilica, Prato della Valle and the hospital district. In particular, the hospital area emerges as a major destination during the morning hours.

The late morning period (9:00--12:00) exhibits a different spatial organization, with increased activity along the east-west corridor containing several tourist attractions, including the Scrovegni Chapel and nearby museums, together with a substantial portion of the university facilities such as departments and libraries. This temporal pattern is compatible with the activation of tourist and student mobility after the main commute period. During the early afternoon, several of the locations that acted as morning destinations become major departure areas, particularly in the central and southern residential neighborhoods of the city.

The present analysis is limited by the relatively small sample size available for the study. However, the same methodology could be applied to substantially larger datasets, potentially exceeding the current sample size by two orders of magnitude, which are available to the mobile phone provider and the collaborating companies; see \cref{sec:data_access}. In particular, a more focused analysis of the city center, combined with a less coarse preprocessing procedure and reduced interference from peripheral mobility behaviors, could provide useful information for policy makers involved in public transport planning, as well as for mobility service providers such as bike-- and scooter--sharing companies interested in optimizing the spatial deployment of their fleets at different times of the day.

\section{Discussion and future work} \label{sec:discussionState}
\noindent
We proposed a state space representation for compositional trajectories and a related algorithm for the model-based clustering of such trajectories, with an application to mobility trajectories in the province of Padova.

A possible improvement involves the creation of the road compositional vectors. We build a circle around the estimated user positions to take into account the uncertainty of the approximation, but in doing so, we use a common radius for all the positions and we do not take into account the different sizes of the coverage areas. We selected a radius of 200 meters based on preliminary findings, which demonstrated that this size is suitable for the dimensions of the cell coverage areas in the city of the case study. However, further research in this direction would be helpful to provide a confidence bound around the estimated positions based on the dimension of the coverage areas and the strength of the phenomenon of bouncing of the telephonic signal in the surroundings of the position.

A Bayesian approach could be considered. The Gaussian state-space model with the Kalman filtering and smoothing procedure has an equivalent Bayesian formulation \citep{durbin2012time}, and the same occurs for simple mixture of experts models \citep{gormley2019mixture}. In model-based clustering, Bayesian nonparametric approaches are usually adopted to model the number of components in the mixture without fixing it in advance \citep{fruhwirth2019handbook}. 
Implementing an MCMC algorithm would be useful for providing posterior distributions and uncertainty estimation for the model parameters. In particular, credible intervals could be provided for the component weight parameters $\boldsymbol{\Gamma}$, as building confidence intervals with the current proposal would require time-consuming resample-based methods. However, issues in Bayesian mixture analysis such as label switching with several clusters would have to be addressed.

\section{Data access and privacy constraints}
\label{sec:data_access}
 The mobile phone data from Fastweb-Vodafone Italia have been handled thanks to the collaboration with Motion Analytica Srl (\url{www.motionanalytica.com}), an Italian innovative company specialized in advanced mobility intelligence. They played a crucial role in the determination of this research project and in its earlier development. The company took care of applying the methods and algorithms we produced to the data in a protected environment, so that we could perform the analysis while respecting the privacy of the phone users. 

Data are irreversibly anonymized, deprived of any personal or sensible data, such as the name, phone number, gender, age, etc, by the mobile network operator provider. Users are only identified via an assigned value to keep track of the same device over time. A further round of preprocessing occurred during the collaboration between the authors and Motion Analytica to ensure that the considered dataset contained only trajectories relative to the actual movement of the users and not purely to the noisy behavior of the telephonic track. The construction of the compositional trajectories described in \cref{sec:data_description} and the computation of the corresponding covariates took place under a similar scenario, having been planned by the authors, who defined the assumptions necessary to obtain adequate data, and then being applied to the technology stack adopted by the company, which handled the operational execution in the production environments.
Finally, device identifiers and other spatio-temporal information were removed from the compositional trajectories. The completely anonymized data were then made available to the authors, who performed the modeling (including the theoretical development) and its application as described in \cref{sec:application}.

\section*{Conflicts of interest}
The authors declare that they have no competing interests.

\section*{Funding}

Andrea Panarotto's research was funded by PON ``Research and Innovation" 2014-2020 Action IV.5 ``PhDs on Green issues", Ministerial Decree 1061/2021, and by the University of Padova -- Department of Statistical Sciences, under the BIRD 2023 funding scheme. Manuela Cattelan acknowledges the financial support provided by the European Union -- Next Generation EU, Mission 4 Component 2 -- CUP C53D23002580006 via the MUR-PRIN grant 2022SMNNKY. Ruggero Bellio was supported by the Departmental Strategic Plan (PSD) of the University of Udine, Department of Economics and Statistics (2022-2025).

\section*{Data availability}
Road-type data were collected from Geofabrik GmbH (\url{www.geofabrik.de}), which provides extracts from the OpenStreetMap project. The data we considered are a subsample of the files available at \url{https://download.geofabrik.de/europe/italy/nord-est.html}.





\section*{Acknowledgments}
The authors thank Denis Cappellari, Bruno Zamengo, and all the Motion Analytica team, as well as Antonio Canale, Sylvia Fr\"uhwirth-Schnatter, Kurt Hornik, and Gianna Serafina Monti for helpful discussion and comments. The authors also thank Fastweb-Vodafone Italia, especially Andrea Zaramella, Dario Di Sorte and Luigi Mancino, for providing fully anonymized data for research purposes.


\bibliography{reference}

\clearpage
\setcounter{page}{1}
\setcounter{equation}{0}
\setcounter{figure}{0}
\setcounter{table}{0}
\setcounter{section}{0}
\renewcommand{\thesection}{\Alph{section}}
\renewcommand{\theequation}{S.\arabic{equation}}
\renewcommand{\thefigure}{S.\arabic{figure}}
\renewcommand{\thetable}{S.\arabic{table}}

\section*{Supplementary Material to ``Model-based Clustering of Compositional Trajectories for the Analysis of Mobility Data''}
\section{Parameter updates}
\label{app:KFS_est_Shumway}
We adopt a notation similar to \cite{shumway2000time}. Let $\mathbf{Z} = \{\mathbf{z}_1, \ldots, \mathbf{z}_n\}$ be a time series of length $n$. Let $\mathbf{a}^{\mathbf{Z}}_t = \mathbb{E}[\mathbf{z}_t\mid \mathbf{z}_1,\ldots,\mathbf{z}_n]$, $\mathbf{P}^{\mathbf{Z}}_t = \mathbb{V}[\mathbf{z}_t\mid \mathbf{z}_1,\ldots,\mathbf{z}_n]$ and \begin{equation*}
    \mathbf{P}^{\mathbf{Z}}_{t,t-1} = \mathbb{E}[(\mathbf{z}_{t} - \mathbb{E}[\mathbf{z}_{t}\mid \mathbf{z}_1,\ldots,\mathbf{z}_t])(\mathbf{z}_{t-1} - \mathbb{E}[\mathbf{z}_{t-1}\mid \mathbf{z}_1,\ldots,\mathbf{z}_t])^\top \mid \mathbf{z}_1,\ldots,\mathbf{z}_n]
\end{equation*}
be, respectively, the smoothed elements, their variances, and the lag-one covariance smoothers. \cite{shumway2000time} update the system matrices as 
\begin{equation*}
    \begin{aligned}
        \TT & = \mathbf{S}_{10}^\mathbf{Z} \left( \mathbf{S}_{00}^\mathbf{Z} \right)^{-1},\\
        \QQ & = n^{-1} \left\{ \mathbf{S}_{11}^\mathbf{Z} - \mathbf{S}_{10}^\mathbf{Z} \left(\mathbf{S}_{00}^\mathbf{Z}\right)^{-1} \left( \mathbf{S}_{10}^\mathbf{Z}\right)^{\top} \right\}, \\
        \HH &=n^{-1} \sum_{t=1}^n\left\{\left(\mathbf{z}_t-\mathbf{a}_t^\mathbf{Z}\right)\left(\mathbf{z}_t-\mathbf{a}_t^\mathbf{Z}\right)^{\top} + \mathbf{P}_t^\mathbf{Z} \right\}\,,
    \end{aligned}
\end{equation*}
where
\begin{equation*}
\begin{aligned}
    \mathbf{S}^{\mathbf{Z}}_{11} & =\sum_{t=1}^n\left(\mathbf{a}_t^n \mathbf{a}_t^{n \top} + \mathbf{P}_t^n\right), \\
    \mathbf{S}^{\mathbf{Z}}_{10} & =\sum_{t=1}^n\left(\mathbf{a}_t^n \mathbf{a}_{t-1}^{n \top} + \mathbf{P}_{t, t-1}^n\right), \\
    \mathbf{S}^{\mathbf{Z}}_{00} & =\sum_{t=1}^n\left(\mathbf{a}_{t-1}^n \mathbf{a}_{t-1}^{n \top} + \mathbf{P}_{t-1}^n\right) .
\end{aligned}
\end{equation*}

We extend their approach to the case where there are multiple sets of system matrices $\boldsymbol{\Theta}^{(k)} = (\TT^{(k)}, \QQ^{(k)}, \HH^{(k)}),\ k= 1,\ldots,K$, multiple time series $\mathbf{Z}^{(i)},\ i = 1,\dots,N$, and each series $\mathbf{Z}^{(i)}$ has length $n_i$ and evolves according to the system matrices $\boldsymbol{\Theta}^{(k)}$ with probability $\hat{w}_{ik}$. Then, the updates for $\TT^{(k)}$ and $\QQ^{(k)}$ become
\begin{equation*}
\begin{aligned}
    \TT^{(k)} &= \mathbf{S}^k_{10} \left(\mathbf{S}^k_{00}\right)^{-1}\,, \\
    \QQ^{(k)}&=\frac{1}{\sum_{i=1}^N \hat{w}_{i k}n_i}\left\{\mathbf{S}^k_{11}-\mathbf{S}^k_{10} \left(\mathbf{S}^k_{00}\right)^{-1} {\mathbf{S}^k_{10}}^{\top}\right\}\,,
\end{aligned}
\end{equation*}
where
\begin{equation*}
    \mathbf{S}^k_{11} = \sum_{i=1}^N \hat{w}_{i k}\mathbf{S}_{11}^{\mathbf{Z}^{(i)}}\,,\quad \mathbf{S}^k_{10} = \sum_{i=1}^N \hat{w}_{i k}\mathbf{S}_{10}^{\mathbf{Z}^{(i)}}\,, \quad \mathbf{S}^k_{00}=\sum_{i=1}^N \hat{w}_{i k}\mathbf{S}_{00}^{\mathbf{Z}^{(i)}}\,.
\end{equation*}
The update for $\HH^{(k)}$ is 
\begin{equation*}
    \HH^{(k)}=\frac{1}{\sum_{i=1}^N \hat{w}_{i k}n_i}\sum_{i=1}^N\sum_{t=1}^{n_i}\hat{w}_{i k}\left\{\left(\mathbf{z}_t-\mathbf{a}_t^{\mathbf{Z}^{(i)}}\right)\left(\mathbf{z}_t-\mathbf{a}_t^{\mathbf{Z}^{(i)}}\right)^{\top} + \mathbf{P}_t^{\mathbf{Z}^{(i)}} \right\}\,.
\end{equation*}

The updates for the component weight parameters in $\boldsymbol{\Gamma}$ are given, at each iteration, by a Newton-Raphson step:
\begin{equation*}
    \boldsymbol{\gamma}_k^{(new)} = \boldsymbol{\gamma}_k^{(old)} -\left\{ \mathcal{H}_k \left( \boldsymbol{\Gamma}^{(old)} \right) \right\}^{-1} \nabla_k Q\left(\boldsymbol{\Gamma}^{(old)}\right)\,, \quad k = 2,\ldots,K\,,
\end{equation*}
where $\nabla_k Q\left(\boldsymbol{\Gamma}\right)$ and $\mathcal{H}_k\left(\boldsymbol{\Gamma}\right)$ are, respectively, the gradient and the Hessian of the $Q$-function in Equation (5) 
with respect to the $k$-th set of component weight parameters, $\boldsymbol{\gamma}_k$. Let $\pi_{ik} = \pi_k\left(\mathbf{x}_i;\, {\mathbf{\Gamma}}\right)$ indicate the component weight parameters defined in \cref{eq:comp_weight_params}. 
The two required elements are obtained explicitly as
\begin{equation*}
    \nabla_k Q \left(\boldsymbol{\Gamma}\right) = \sum_{i=1}^N \left(\hat{w}_{i k} - \pi_{ik}\right) \tilde{\mathbf{x}}_{i}\,,\quad    \mathcal{H}_k\left(\boldsymbol{\Gamma}\right) = - \sum_{i=1}^N \pi_{ik} (1-\pi_{ik}) \tilde{\mathbf{x}}_{i} \tilde{\mathbf{x}}_{i}^\top\,.
\end{equation*}

\section{Simulation study}
A simulation study is carried out to verify the clustering performance of the model. We repeat the experiment 100 times, each time generating 3 groups of compositional trajectories, with simplex dimension $D=4$. The sizes of the groups are sampled each time between 20 and 50, and each series length is sampled uniformly between 10 and 30. We suppose that the evolution of the compositions depends on their group. In particular, given a starting composition $\mathbf{y}_{i0}$ uniformly sampled in the simplex $\mathbb{S}^3$, the following simulation scheme is implemented.
\begin{itemize}
    \item The series in Group 1 evolve according to the perturbation
    \begin{equation*}
        \mathbf{y}_{i,\,t+1} = \mathbf{y}_{it}\oplus    \mathbf{a} \oplus \mathbf{u}_{it},    
    \end{equation*}
    where $\mathbf{a} = (0.5,0.2,0.2,0.1)$ is a fixed composition and the $\mathbf{u}_{it}$ play the role of random errors and are sampled from a Dirichlet distribution $\mathcal{D}(\boldsymbol{\alpha}),\ \alpha_j = 1$ for $j=1,\ldots,D$.
    \item The series in Group 2 evolve according to the perturbation $\mathbf{y}_{i,\,t+1} = \mathbf{y}_{it} \oplus \mathbf{u}_{it}$, with $\mathbf{u}_{it}\sim \mathcal{D}(\boldsymbol{\alpha}),\ \alpha_j = 2$ for $j=1,\ldots,D$. The perturbation by a fixed composition is dropped.
    \item The series in Group 3 completely drop the dependence from the previous time points, and $\mathbf{y}_{it}$ is uniformly sampled in the simplex for all $t = 1, \ldots, n_i$.
\end{itemize}
After creating the series, a covariate $x_i,\ i= 1,\ldots,N$, is sampled for each series to verify whether the mixture of experts model helps the clustering procedure. Hence, $x_i$ is sampled from a $\mathcal{N}(-1,1)$ for the series of the first group, from a standard normal distribution for the series of the second group, and from a $\mathcal{N}(1,1)$ for the series of the third. We run the model twice, including or excluding the covariate $x$ in the analysis. The number of components $K$ varies between 2 and 8. Clusters are initialized using \texttt{dtwclust} \citep{sarda2019time}.

\begin{figure}[tp]
    \centering
    \begin{tikzpicture}
        \small
        \node[] at (0,0) (table)
        {
        \begin{tabular}{c|c}
            $\hat{K}$ & Freq. (Int/Cov)  \\
            \midrule
            2 & 1 / 3 \\
            3 & 81 / 80 \\
            4 & 16 / 16 \\
            5 & 2 / 1 \\
        \end{tabular}};
        \node[right = 1em of table] (boxplot) {\includegraphics[width=0.75\linewidth]{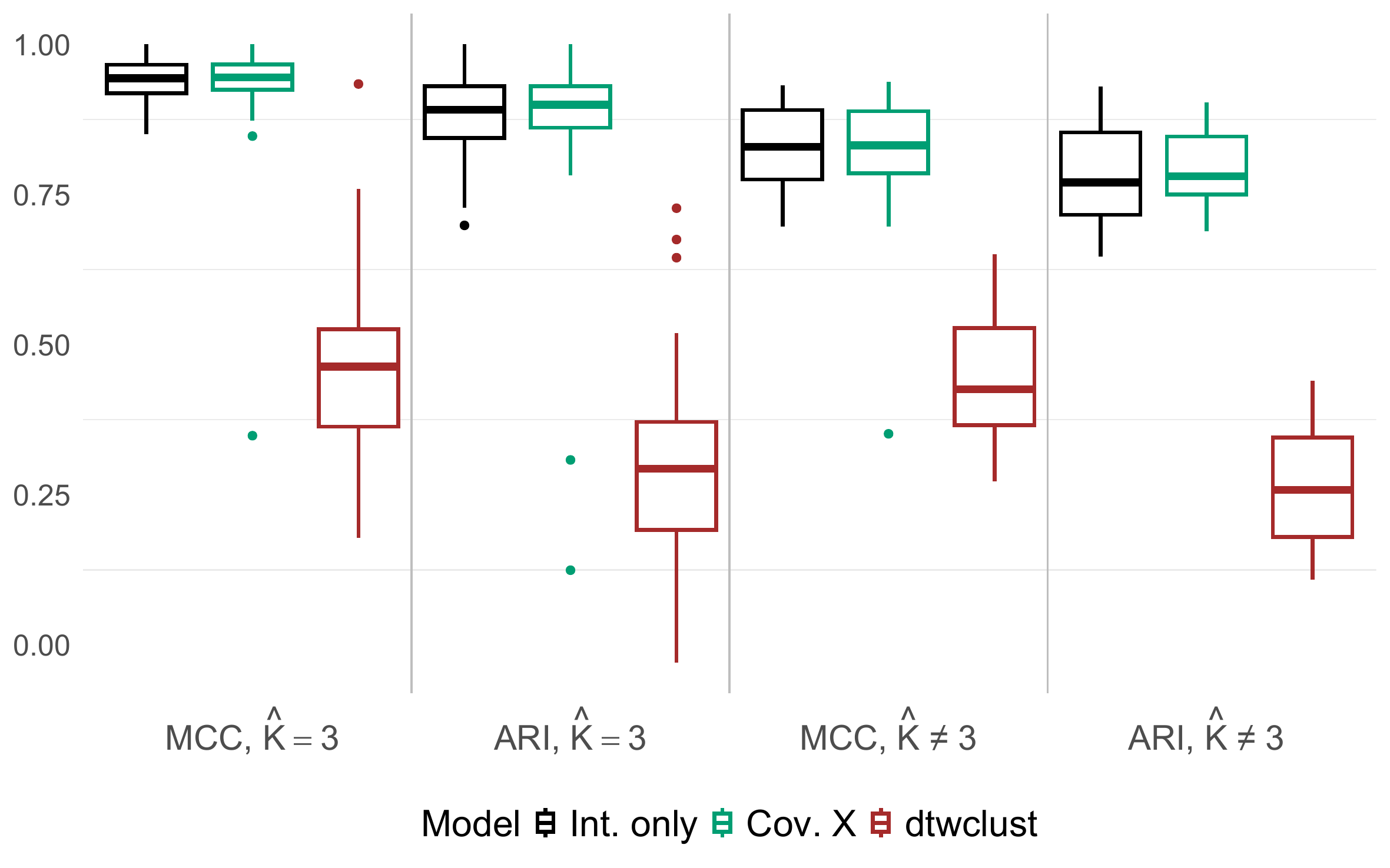}};
    \end{tikzpicture}
    \caption{Results in the selection of the number of components and boxplot of the MCC and ARI scores of the assignment of the series to the clusters across the simulation experiments. We compare the proposed model excluding (Int.~only) or including (Cov.~X) the cluster-specific distributed covariate $X$, with the clustering based on DTW distance (\texttt{dtwclust}).
    We distinguish the cases in which $K$ is correctly estimated $(\hat{K}=3)$ from the cases where it is not $(\hat{K}\neq3)$.
    }
    \label{fig:mcc_scores}
\end{figure}

\cref{fig:mcc_scores} shows the results in the selection of the number of components $\hat{K}$ and in the allocation of the series to the clusters. The table shows the estimated $\hat{K}$ across the 100 experiments, according to the ICL-BIC criterion in \cref{eq:ICL-BIC_penalization}. The correct number of components, $\hat{K} = 3$ is the most selected by both the model that includes the covariate and the one that excludes it.

To verify whether the series are allocated to the correct cluster, we use, along with the Adjusted Rand Index \citep[ARI;][]{rand_objective_1971,hubert_comparing_1985}, the multi-class Matthews correlation coefficient \citep[MCC;][]{matthews_comparison_1975,gorodkin_comparing_2004}, an extension of the Pearson correlation coefficient for contingency tables, defined as
\begin{equation*}
    {\rm MCC}  = \frac{c N - \mathbf{t}\cdot\mathbf{p}}{\sqrt{N^2 - \mathbf{p}\cdot\mathbf{p}}\ \sqrt{N^2 - \mathbf{t}\cdot\mathbf{t}}}\,.
\end{equation*}
The value $c$ represents the number of correctly allocated series. The vector $\mathbf{t} = (t_1,\ldots,t_K)$ has each element $t_k$ that corresponds to the true size of the group $k$, while the vector $\mathbf{p} = (p_1,\ldots,p_{\hat{K}})$ has each element $p_k$ corresponding to the count of elements allocated by the model in cluster $k$. When $\hat{K}\neq K$, the scalar product $\mathbf{t}\cdot\mathbf{p}$ is intended as $\sum_{k=1}^{K^\prime} t_kp_k$, where $K^\prime = \min(K,\hat{K})$. To account for label switching, all possible permutations of the labels are considered, and the maximum MCC score is selected.

The boxplots in \cref{fig:mcc_scores} show the values of the ARI and MCC scores of the labels predicted by the models that exclude or include the covariate and by the method in package \texttt{dtwclust}, based on the DTW distance, and used as initialization for both this simulation and the application in \cref{sec:application}. We separate the cases in which the number of clusters is correctly estimated ($\hat{K}=3)$ from the cases where it is not.

\begin{figure}[tp]
\centering

\begin{subfigure}{0.8\linewidth}
\centering
\includegraphics[width=\linewidth]{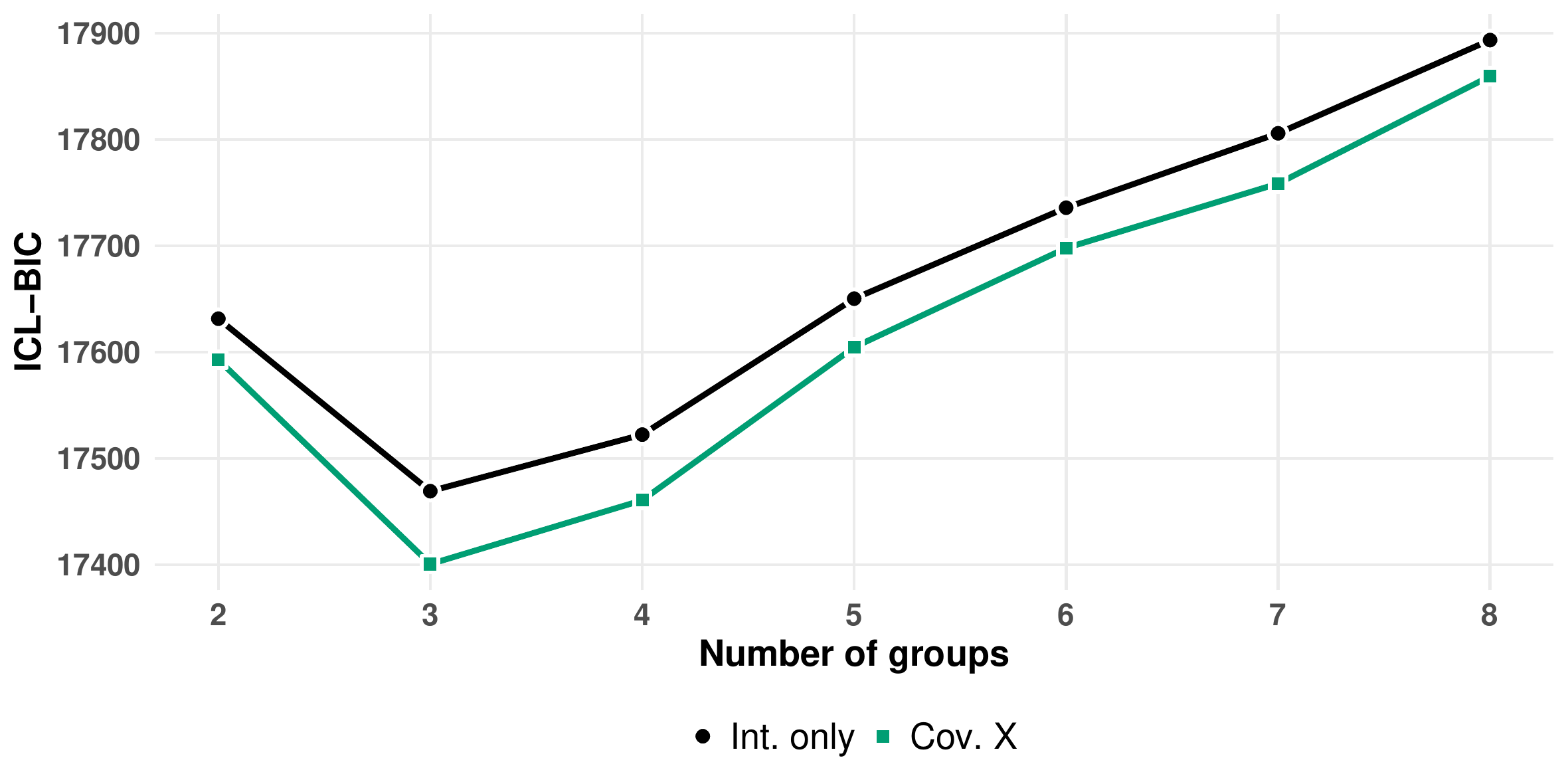}
\caption{ICL-BIC for the simulated data as the number of components changes. The minimum is for $K=3$.}
\label{fig:BIC_simulation}
\end{subfigure}

\vspace{1em}

\begin{subfigure}{0.9\linewidth}
\centering
\includegraphics[width=\linewidth]{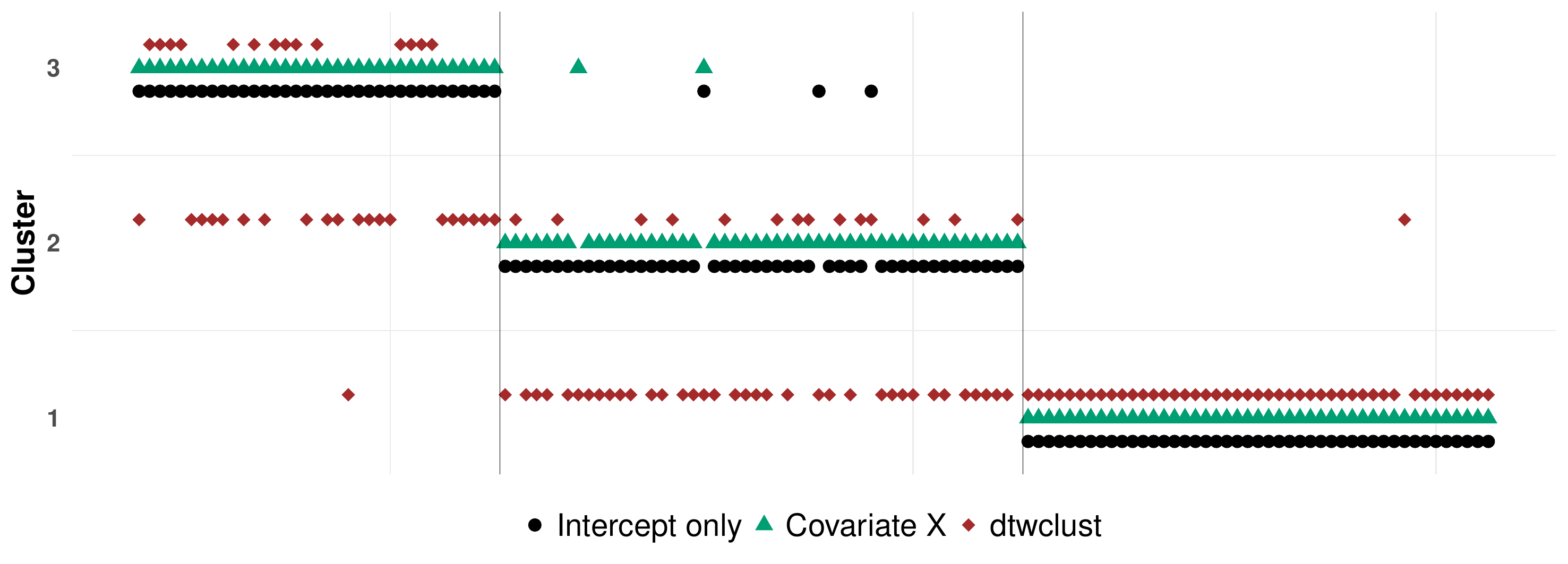}
\caption{Allocation of the simulated series. The dots indicate the cluster to which each series is allocated according to the model. The black vertical lines separate the true groups.}
\label{fig:allocation_simulation}
\end{subfigure}

\caption{Results in the first simulation experiment. The proposed models are indicated by ``Intercept only'' and ``Covariate X'' according to the inclusion of the component-specific covariate in the mixture of experts, and are compared with the clustering based on DTW distance.}
\end{figure}

The two state-space models show similar results and outperform distance-based clustering, with large values of both metrics that indicate that the series allocation in the groups is mostly correct, even when the estimated number of clusters is wrong. When $\hat{K}=3$, ARI and MCC scores are close to 1, indicating an almost perfect allocation when the correct number of components is retrieved. The presence of the covariate slightly improves the allocation results, but since $x$ does not interfere with the initial assignment of the series to the groups and with their creation, the model based only on the compositional trajectory shows satisfactory performance. However, the presence of the group-specific covariate generally improves the complete likelihood, to underscore that it effectively helps with the allocation probabilities of some series. This effect is illustrated in \cref{fig:BIC_simulation}, which depicts the ICL-BIC scores of the models in one of the simulation experiments. In addition, \cref{fig:allocation_simulation} shows how the series are allocated to the clusters in the same experiment. Up to label switching, the models based on the state-space representation are able to collocate the series in the clusters correctly, with just two series that are misplaced when the covariate is included in the model and three when it is not included. Clustering based on DTW distance is not able to retrieve the correct clusters in this setting.

\section{Conditional backward elimination procedure}
\label{app:cond_elimination}
\noindent
During the model selection phase, the analysis has to be performed multiple times, while varying the number of components, to find the best number of clusters. However, computing every model from scratch may not lead to satisfactory results.
The ICL-BIC curve may become unstable as the number of components varies. When $K$ is low, this variability is more evident, probably due to the reach of a local minimum. When $K$ is large, overparametrization allows the assignment of the series to a tailored cluster, so the risk of ending with a local minimum is lower.

In light of this, the following proposal exploits the better fit of the high-components cases to help the analysis in the low-components cases, while reducing the above-mentioned overparametrization. The algorithm is started with a high number of components, e.g. 12, and after convergence, one component is removed, and the computation restarts with the results from the remaining components as initial parameters. 

\begin{figure}[htp]
    \centering
    \includegraphics[width=0.7\linewidth, trim = 0 8cm 0 2cm, clip]{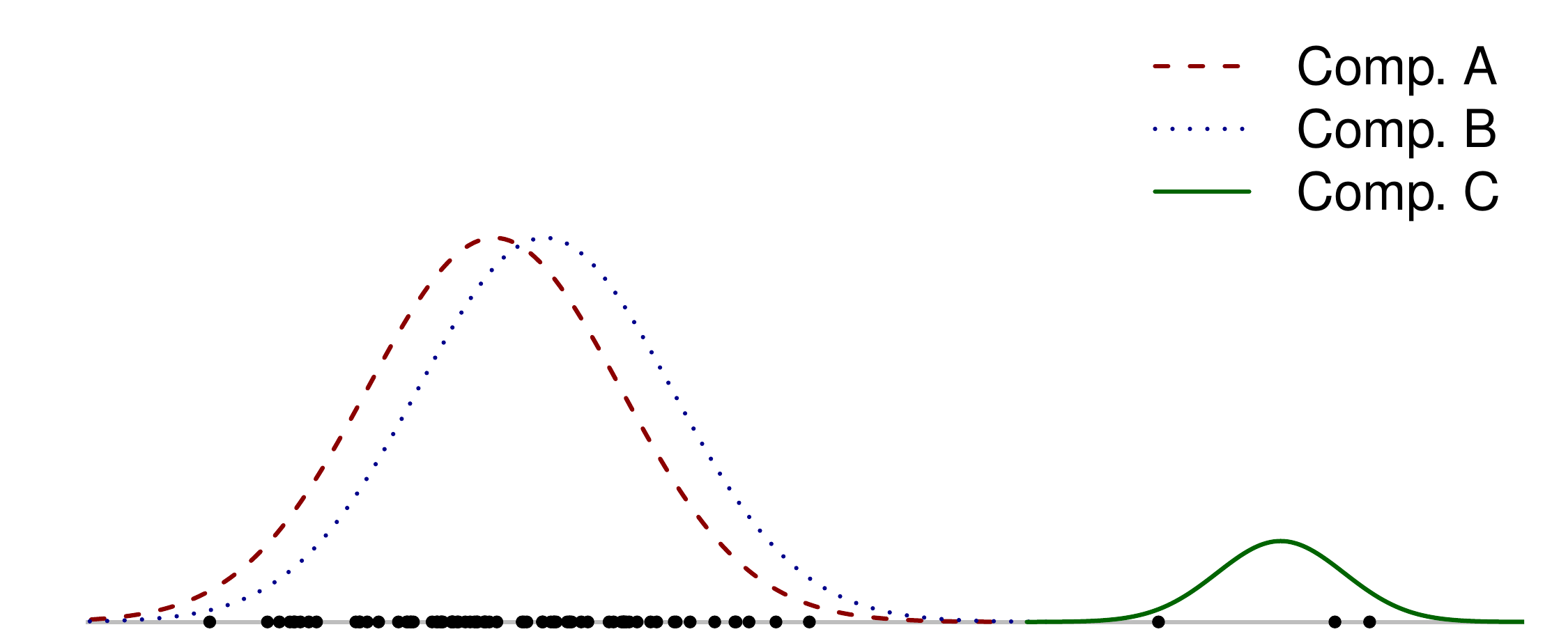}
    \caption{Mixture model toy example.}
    \label{fig:mix_example}
\end{figure}

The choice of the component to remove is crucial. \cref{fig:mix_example} pictures a toy example to visualize the underlying idea. Assume that points are sampled from three Gaussian distributions: two components, represented respectively by the dashed and dotted lines, have a large weight in the distribution and are almost overlapping, and a smaller component, far away from the others, represented by a continuous line.

If one wants to reduce the number of components, a sound strategy is to merge the two overlapping clusters into one and keep the smaller cluster separate. To achieve this, note that after a clustering procedure, the cluster represented by the continuous line would have just a few points assigned, but each of them would have a large probability of belonging to it. The clusters corresponding to the dashed- and dotted-line components would have more assigned points each, but these points could also belong to the other cluster with considerable probability. Our procedure captures this idea and applies it to the trajectory clustering by first assigning the series to the clusters using the estimated latent group membership indicators, and, conditional on the assignment, calculating the average probability of the points to belong to the component.

Let $S_k=\left\{i: \hat{C}_i = k\right\}$, for $k=1,\ldots,K$, be the sets of all the indices relative to the units assigned to each cluster. The conditional average weight is defined as 
\begin{equation}
    \tilde{\eta}_k=\frac{1}{\# S_k} \sum_{i \in S_k} \hat{w}_{i k}\,.
    \label{eq:example_conditional}
\end{equation}
For each cluster, this value ranges between $K^{-1}$, in the extreme case where all components overlap exactly, and 1, if the corresponding component is well separated from the rest of the mixture. When $\tilde{\eta}_k$ is low, the elements in the cluster have low probability of belonging to the corresponding component and could quite likely be part of another one. In the end, we remove the component $k$ with the smallest value for $\tilde{\eta}_{k}$.

\begin{figure}[htp]
    \centering
    \includegraphics[width=0.7\linewidth, trim = 0 8cm 0 2cm, clip]{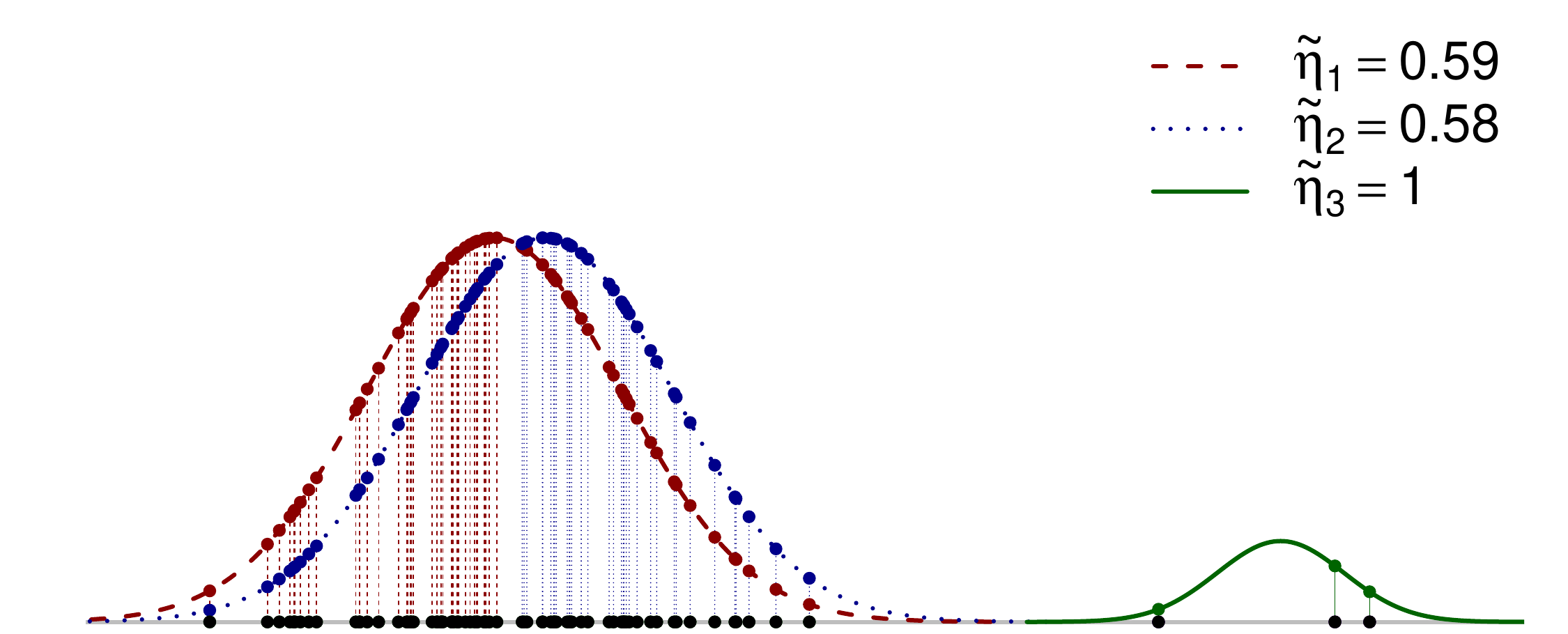}
    \caption[Conditional average weights for the conditional backward elimination procedure]{Conditional average weights in the sense of \cref{eq:example_conditional} for the conditional backward elimination procedure.}
    \label{fig:mix_conditional}
\end{figure}

\cref{fig:mix_conditional} shows the values of the conditional average weights for the components in the example. The dotted-line component would be removed, and, at the next iteration, all of its elements would be assigned to the dashed-line component. We call this heuristic process `conditional backward elimination procedure', since the calculation of the score to eliminate the component is done conditionally on the group assignment. 

\cref{subfig:BicWD}, in the main text, shows the results of model selection for the case study application, starting with 12 components and using the conditional backward elimination procedure.

\section{Further details in cluster analysis}
\label{supp:altre_mappe}
\begin{figure}[tp]
\centering

\begin{subfigure}{0.55\linewidth}
\centering
\includegraphics[width=\linewidth]{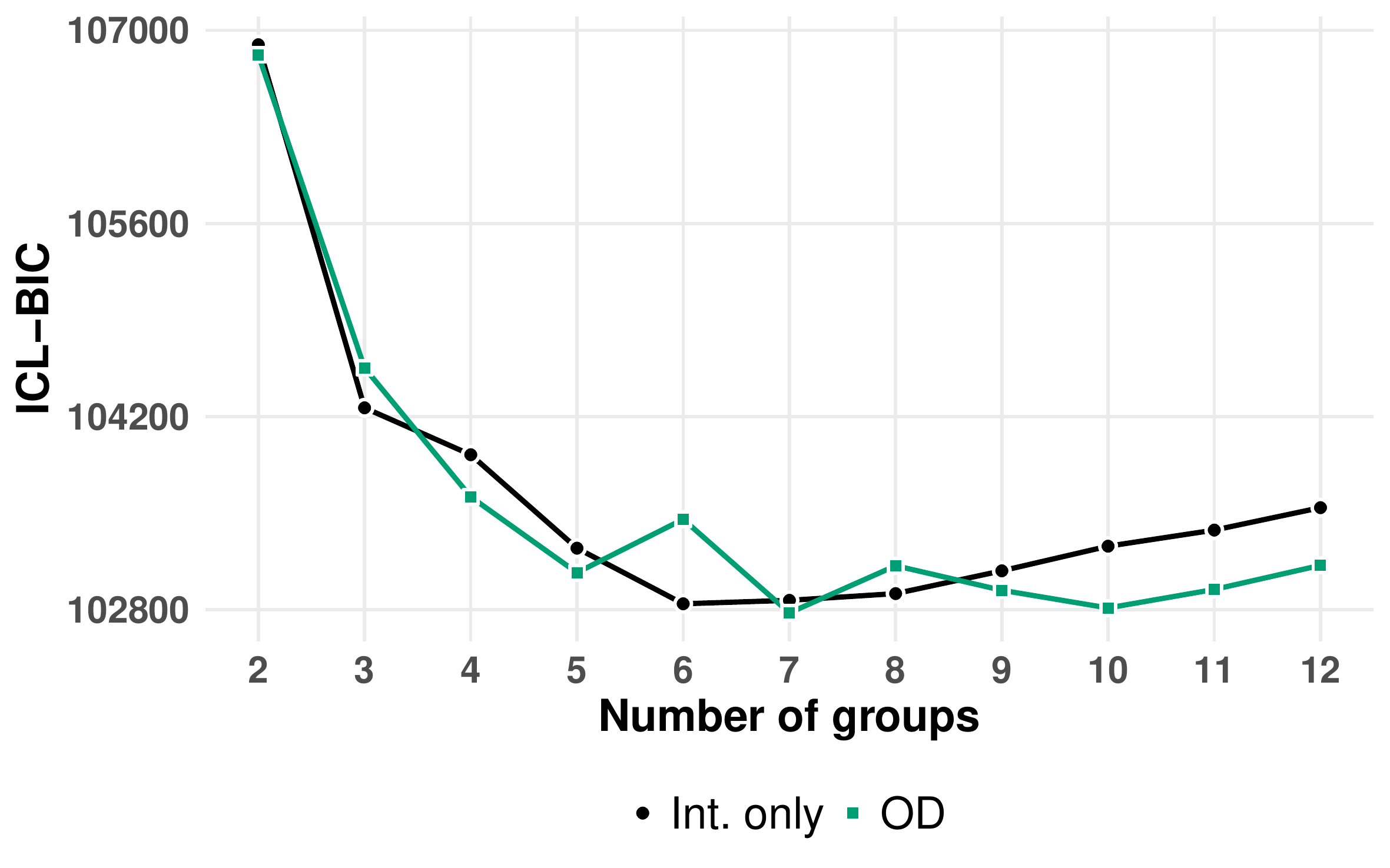}
\caption{Model selection based on ICL-BIC.}
\label{subfig:BicWE}
\end{subfigure}
\hfill
\begin{subfigure}{0.4\linewidth}
\centering
\includegraphics[width=\linewidth]{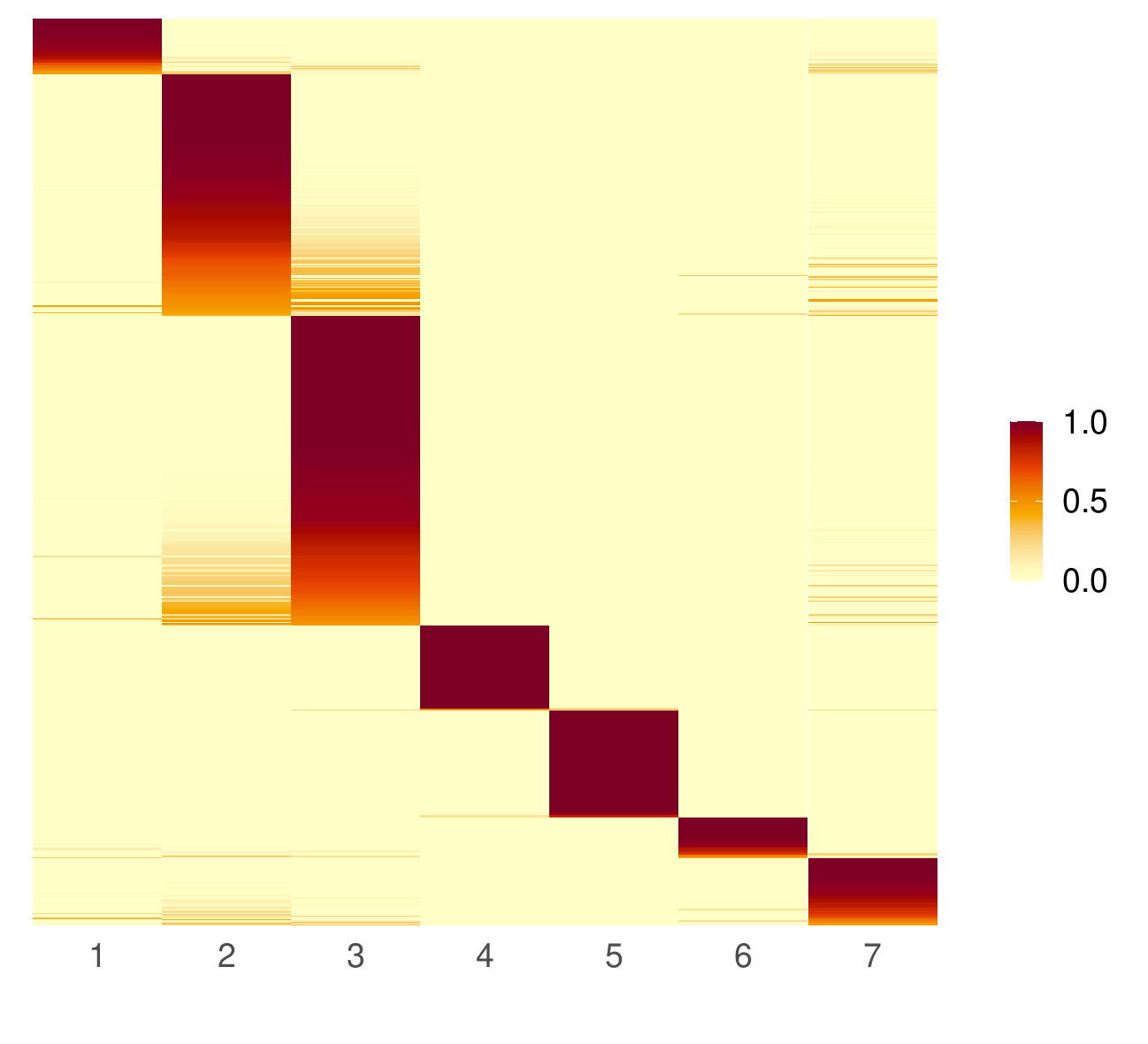}

\caption{Assignment probabilities.}
\label{subfig:assign_heatmap_WE}
\end{subfigure}

\caption{Weekend analysis model selection and estimated assignment probabilities ($\hat{w}_{ik}$) to the components (on the columns) for the weekend series (on the rows).}
\label{fig:result_WE}
\end{figure}

\cref{fig:result_WE} shows the selection of the model for the weekend analysis.  The best model in terms of ICL-BIC includes the OD distance as a trajectory covariate and selects $\hat{K} = 7$ components. The heatmap of the assignment probabilities indicates again a good separation between the clusters, with some exchange only between the largest clusters, 2 and 3, containing trajectories on the highway and the ring road. 

\begin{figure}[htp]
\centering
\begin{subfigure}{0.47\linewidth}
\centering
\includegraphics[width=\linewidth]{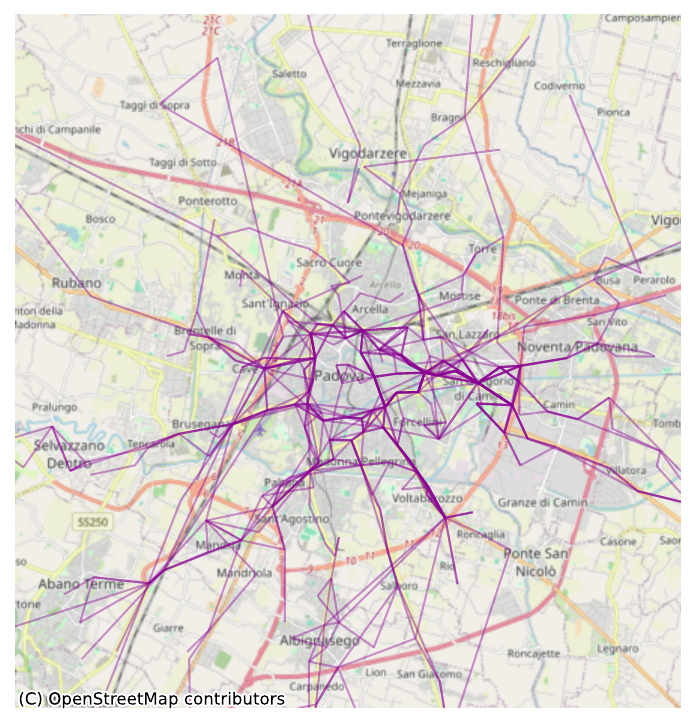}
\caption{Cluster 3, 87 trajectories, $\hat{\gamma}_{3,OD}=-0.86$.}
\label{fig:map_cl3}
\end{subfigure}
\hfill
\begin{subfigure}{0.47\linewidth}
\centering
\includegraphics[width=\linewidth]{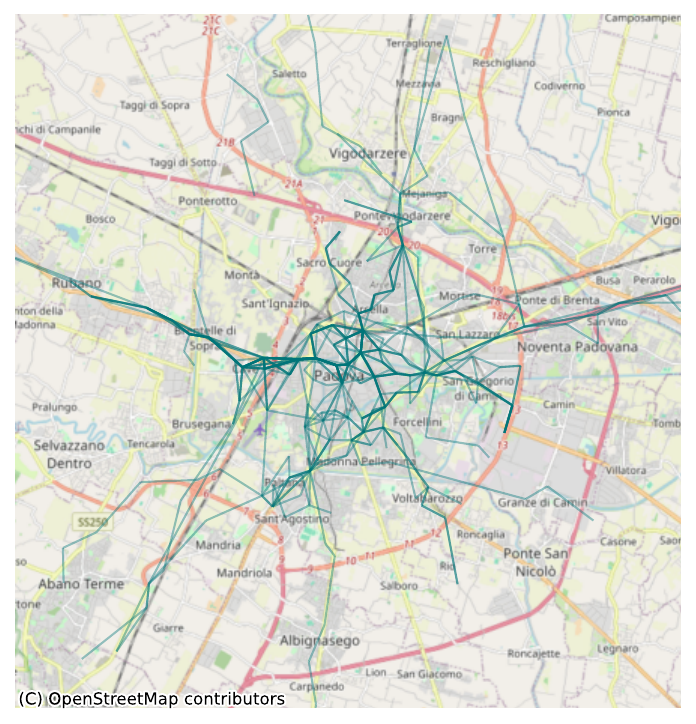}
\caption{Cluster 5, 75 trajectories, $\hat{\gamma}_{5,OD}=-1.41$.}
\label{fig:map_cl5}
\end{subfigure}

\medskip

\begin{subfigure}{0.47\linewidth}
\centering
\includegraphics[width=\linewidth]{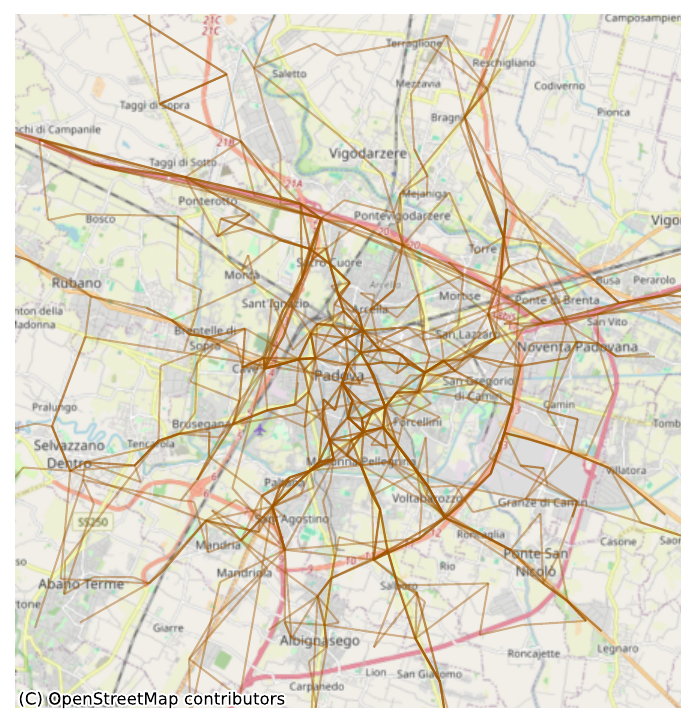}
\caption{Cluster 6, 110 trajectories, $\hat{\gamma}_{6,OD}=-0.30$.}
\label{fig:map_cl6}
\end{subfigure}
\hfill
\begin{subfigure}{0.47\linewidth}
\centering
\includegraphics[width=\linewidth]{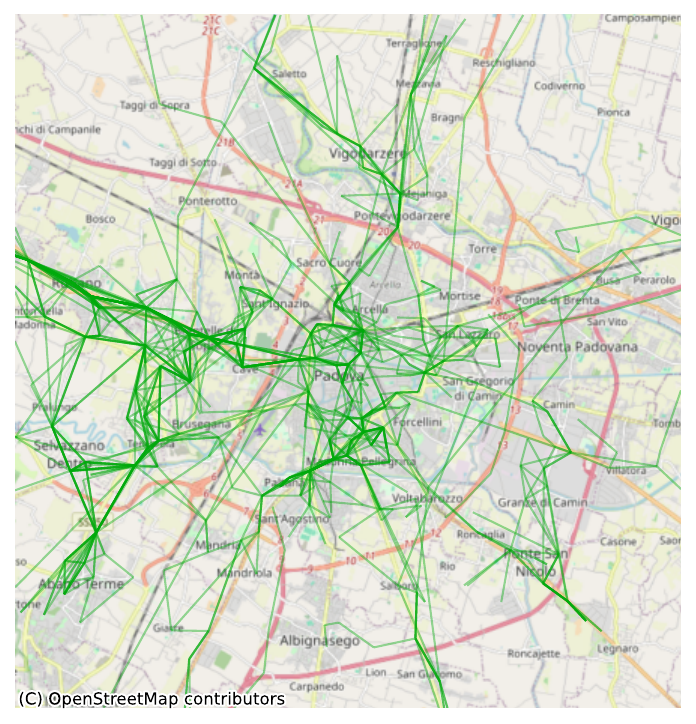}
\caption{Cluster 8, 190 trajectories, $\hat{\gamma}_{8,OD}=-1.11$.}
\label{fig:map_cl8}
\end{subfigure}

\caption{Trajectory clusters plotted on map for the remaining clusters for the weekday analysis.}
\label{fig:map_cl_supp}
\end{figure}

\begin{figure}
\begin{subfigure}{0.47\linewidth}
\centering
\includegraphics[width=\linewidth]{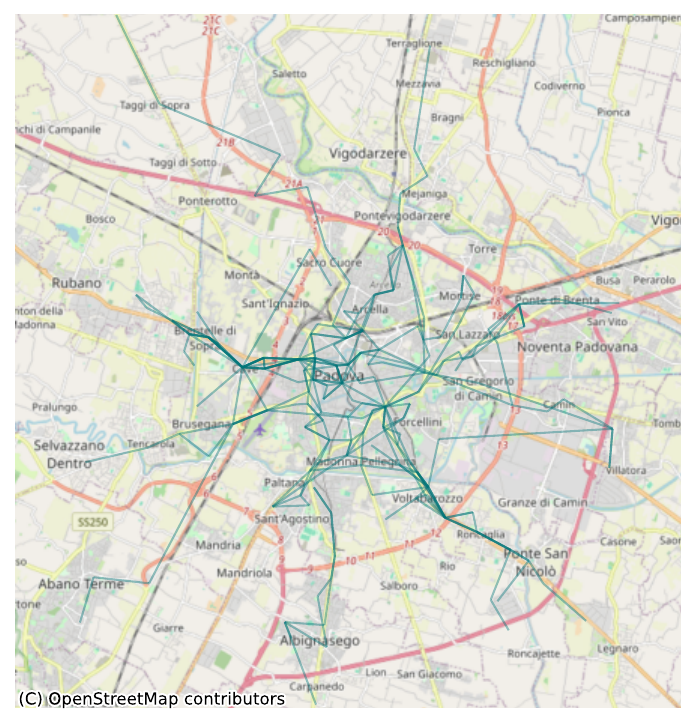}
\caption{Cluster 1, 44 trajectories, $\hat{\gamma}_{1,OD}=-1.4$.}
\label{fig:map_cl_1WE}
\end{subfigure}
\hfill
\begin{subfigure}{0.47\linewidth}
\centering
\includegraphics[width=\linewidth]{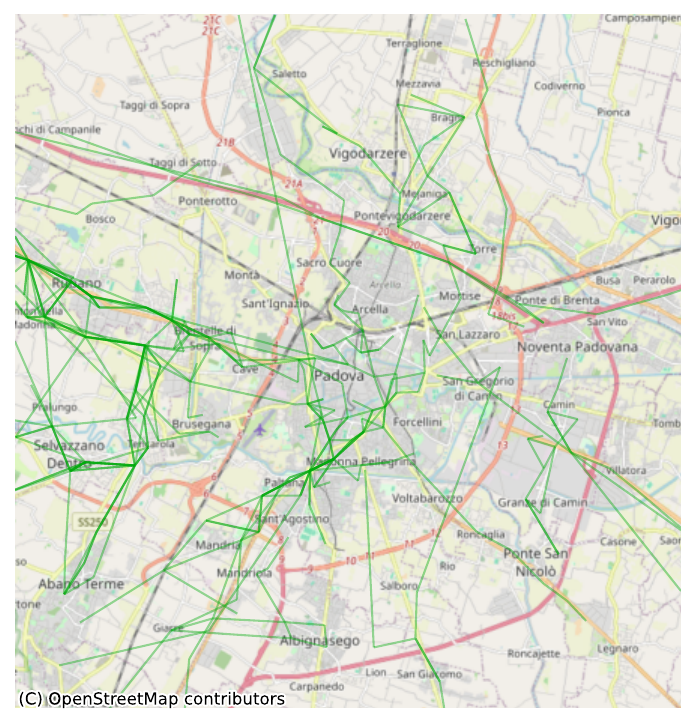}
\caption{Cluster 4, 67 trajectories, $\hat{\gamma}_{4,OD}=-0.77$.}
\label{fig:map_cl_4WE}
\end{subfigure}

\medskip

\begin{subfigure}{0.47\linewidth}
\centering
\includegraphics[width=\linewidth]{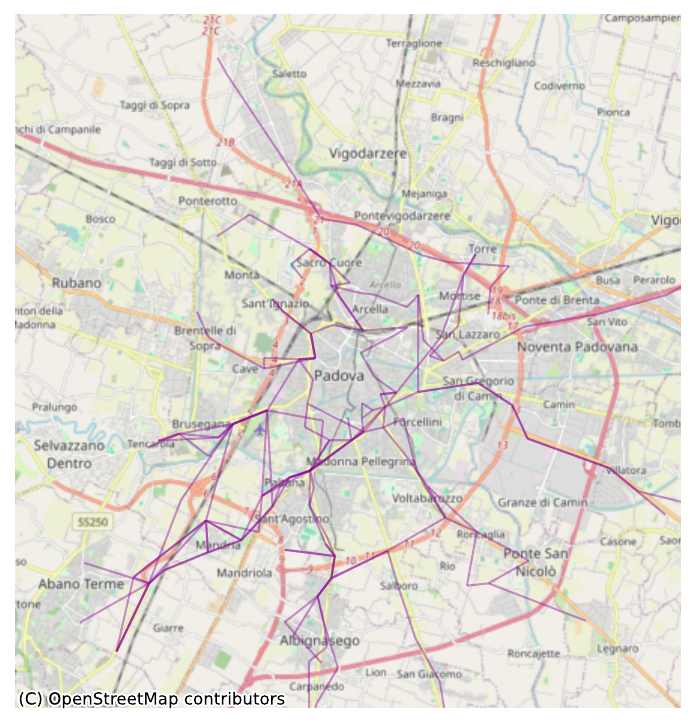}
\caption{Cluster 6, 32 trajectories, $\hat{\gamma}_{6,OD}=-1.30$.}
\label{fig:map_cl_6WE}
\end{subfigure}
\hfill

\caption{Trajectory clusters plotted on map for the weekend analysis.}
\label{fig:map_cl_supp_WE_57}
\end{figure}

\cref{fig:map_cl_supp,fig:map_cl_supp_WE_57} represent the trajectories of the clusters that were not present in \cref{fig:map_cl,fig:map_cl_WE_14} of the main text, respectively, for the weekday and weekend analysis. Also in this case, some similarities may be found, especially between clusters 1, 4, and 6 on the weekend and 5, 8, and 3 on weekdays, respectively, but the small cardinality could be misleading in finding the correct associations.
Each cluster presents its own peculiarities that could be relevant to policy makers in a detailed mobility analysis.


\end{document}